\documentclass[10pt,journal,compsoc]{IEEEtran}
\ifCLASSOPTIONcompsoc
  \usepackage[nocompress]{cite}
\else
  \usepackage{cite}
\fi
\ifCLASSINFOpdf
   \usepackage[pdftex]{graphicx}
   \graphicspath{{./figures/}}
    \DeclareGraphicsExtensions{.jpg,.jpeg,.pdf,.png} 
\else
\fi
\usepackage{dsfont}
\usepackage{amsmath}
\usepackage{amssymb}
\usepackage{bbm}
 \usepackage[caption=false,font=footnotesize,labelfont=sf,textfont=sf]{subfig}

\usepackage[usenames, dvipsnames]{color}
\usepackage[dvipsnames]{xcolor}
\usepackage[hidelinks]{hyperref}
\usepackage{enumitem}

\usepackage{soul}

\usepackage{setspace}
\usepackage{tabu} 

\begin{document}
%
\title{A Deep Generative Model for \\Reordering Adjacency Matrices}
%
%
%
%

\author{Oh-Hyun~Kwon,
        Chiun-How Kao,
        Chun-houh~Chen,
        and~Kwan-Liu~Ma
\IEEEcompsocitemizethanks{\IEEEcompsocthanksitem O. Kwon is with the Department of Computer Science, University of California, Davis, Davis, CA, 95616. E-mail: kw@ucdavis.edu
\IEEEcompsocthanksitem C. Kao is with the Department of Statistics, Tamkang University, New Taipei City 251301, Taiwan. Email: 157294@mail.tku.edu.tw
\IEEEcompsocthanksitem C. Chen is with the Institute of Statistical Science, Academia Sinica, Taipei 11529, Taiwan. E-mail: cchen@stat.sinica.edu.tw
\IEEEcompsocthanksitem K. Ma is with the Department of Computer Science, University of California, Davis, Davis, CA, 95616. E-mail: ma@cs.ucdavis.edu}
}
\IEEEtitleabstractindextext{%
\begin{abstract}
Depending on the node ordering, an adjacency matrix can highlight distinct characteristics of a graph. 
Deriving a ``proper'' node ordering is thus a critical step in visualizing a graph as an adjacency matrix.
Users often try multiple matrix reorderings using different methods until they find one that meets the analysis goal.
However, this trial-and-error approach is laborious and disorganized, which is especially challenging for novices.
This paper presents a technique that enables users to effortlessly find a matrix reordering they want.
Specifically, we design a generative model that learns a latent space of diverse matrix reorderings of the given graph.
We also construct an intuitive user interface from the learned latent space by creating a map of various matrix reorderings.
We demonstrate our approach through quantitative and qualitative evaluations of the generated reorderings and learned latent spaces. 
The results show that our model is capable of learning a latent space of diverse matrix reorderings.
Most existing research in this area generally focused on developing algorithms that can compute ``better'' matrix reorderings for particular circumstances.
This paper introduces a fundamentally new approach to matrix visualization of a graph, where a machine learning model learns to generate diverse matrix reorderings of a graph.
\end{abstract}

}

\maketitle

\IEEEdisplaynontitleabstractindextext

%
\IEEEpeerreviewmaketitle

\IEEEraisesectionheading{\section{Introduction}\label{sec:introduction}}

\IEEEPARstart{G}{raph-structured} data are widely found in many disciplines; examples include protein-protein interaction networks in biological science, brain networks in neuroscience , and friendship networks in social science.
Visualizing a graph allows people to intuitively understand and communicate the structural information in the data, which is especially helpful for non-experts in network analysis.
A popular way to visualize a graph is visually encoding an adjacency matrix.

The rows and the columns of an adjacency matrix correspond to the nodes of a graph, where a matrix element $A_{i,j}$ indicates if there is an edge between node $u_i$ and node $u_j$.
A view of an adjacency matrix is generally made by coloring an area mark for each element of the matrix based on its value \cite{Munzner14:VAD}.
Although node-link diagrams are more commonly used for visualizing graphs, adjacency matrix views are often preferred for dense graphs.
Matrix views do not suffer from occlusion issues of node-link diagrams, such as node overlaps and edge crossings.
Thus, adjacency matrix views can have higher perceptual scalability and information densities than node-link diagrams \cite{Munzner14:VAD}.

A node-link diagram of a graph can be made to display distinct structural characteristics of the graph depending on how the nodes are laid out in a two- or three-dimensional space.
In an adjacency matrix view, nodes are laid out linearly in terms of their order along the rows (and the columns) of the matrix; and different node orderings often highlight different structural aspects of the same graph (e.g., \autoref{dmr:fig:comparison-reord-methods}).
Therefore, it is essential to find those ``good'' node orderings that show the aspects of a graph that users wish to examine or present.

Decades of research have introduced many matrix reordering methods, which sort the rows and the columns of a matrix to reveal the structural information in the data \cite{WuTzengChen08, Liiv10, Behrisch16}.
However, there is no ``best'' method, as each method follows a different set of heuristics that can reveal certain structures in a graph.
Therefore, users typically try multiple reorderings of the given adjacency matrix using different methods until they find one that best meets a specific visualization goal.
Unfortunately, this trial-and-error approach is time-consuming and haphazard, which is especially challenging for novices.

This paper presents a technique that builds an intuitive interface for users to effortlessly find the desired adjacency matrix views of a graph.
The fundamental approach is similar to the one for node-link diagrams presented by Kwon and Ma \cite{Kwon19}.
We first collect a set of reordering examples of the adjacency matrix of the given graph.
Then, we train a generative model that learns a latent space of diverse matrix reorderings of the given graph.
Finally, based on a grid of samples from the learned latent space, we build a what-you-see-is-what-you-get (WYSIWYG) interface that enables users to intuitively depict diverse reorderings of the given adjacency matrix.

However, reordering an adjacency matrix is a discrete problem, while laying out a node-link diagram is a continuous problem.
Thus, designing a neural network for matrix reordering faces unique challenges.
Overall, we identify three key challenges and design a novel neural network architecture addressing said challenges:
\begin{itemize}[topsep=0pt,itemsep=-1ex,partopsep=1ex,parsep=1ex,leftmargin=16pt]
\item Slight differences in node positions are negligible in a node-link diagram.
In the case of an adjacency matrix, however, a single difference may no longer represent the same graph.
To ensure the generated adjacency matrices represent the same structure as the input graph, we train the model to produce permutation matrices, which represent node orderings, and then permute the given adjacency matrix, not to approximate reordered adjacency matrices naively.
\item Gradient-based optimization is the standard approach to training a neural network \cite{DeepLearningNature}. 
However, it is only applicable to differentiable functions.
Since reordering a matrix is not a differentiable problem, we need a differentiable approximation.
We employ a relaxed sorting operator (e.g., Sinkhorn~\cite{Sinkhorn} and SoftSort~\cite{SoftSort}) such that standard gradient-based optimization techniques can be used for the training.
\item Depending on the structure of the given graph, some distinct node orderings may produce the same matrix reordering. 
Training a model to learn different solutions for the same output could lead to ineffective or inefficient results.
We thus design the neural network architecture and the loss function to take the resulting reordered matrices into account but not how the nodes are ordered (permutation equivariant).
\end{itemize}
Our evaluation results show that the presented architecture is capable of learning a generative model that can visualize a graph in diverse matrix reorderings that are comprehensive and distinctive. 
This new capability will greatly aid those who rely on matrix visualization extensively in their study of complex network data.

\section{Related Work}
Designing a deep generative model for matrix reordering is related to matrix reordering methods, deep learning for sorting, and deep generative models.

\subsection{Matrix Reordering}
\label{dmr:sec:related-work:reord}
Matrix reordering is a problem of finding a ``good'' ordering of the rows and columns of a relational matrix (e.g., an adjacency matrix and a correlation matrix) to reveal structural information from a set of objects.
This problem is also known as (or equivalent to) \emph{seriation}, \emph{linear arrangement}, and \emph{sequencing}.
Since there can be $O(n!)$ different permutations of a set of $n$ objects, a brute-force enumeration approach to finding a good node ordering is infeasible even for a graph with a few dozens of nodes.

Decades of research across various disciplines have introduced many methods for matrix reordering, where each method uses a different set of heuristics, such as traveling salesman problem solvers \cite{TSPOrdering}, spectral ordering (i.e., the order of the Fiedler vector) \cite{SpectralOrdering, DingHe04}, and hierarchical clustering \cite{Eisen98, OLO, Gruvaeus72}.
Several studies \cite{WuTzengChen08, Liiv10, Behrisch16, Hahsler17} have reviewed various matrix reordering methods and have shown that different methods can show distinct patterns in the same matrix.

Unfortunately, it is not clear how to derive a suitable reordering for the given dataset and analysis goal \cite{Behrisch17, GUIRO}.
Users typically first produce multiple reorderings of the given matrix using various methods and then select one that fits the given circumstances.
Several techniques have been introduced to support users through such a trial-and-error process by providing interactive methods to steer matrix reordering algorithms \cite{MatrixExplorer, PermutMatrix, COMBat, GUIRO}.
The goal of these techniques is to narrow down the search space.
Therefore, multiple trials with continuous human intervention would be required to explore other possible reorderings of the same matrix.
In contrast, our approach is to build a model that can produce diverse reorderings of the given matrix, not to narrow the search space.
Moreover, we train such a model in a fully unsupervised manner, thus not requiring human intervention.

\subsection{Deep Generative Models}
Deep generative models, such as variational autoencoders \cite{VAE} and generative adversarial networks \cite{GAN}, have shown great success in various domains, such as image generation \cite{VQVAE2, BigGAN}, text generation \cite{Guu18TextGen, Zhang17TextGen}, and music generation \cite{Jukebox, GANSynth}.

In the field of data visualization, Kwon and Ma \cite{Kwon19} recently introduced a deep generative model for visualizing a graph in node-link diagrams.
Their approach is to learn a two-dimensional latent space of diverse layouts of a graph.
Then, users use the learned latent space as a map of various node-link diagrams of the given graph, such that they can effortlessly find a layout they want.

Inspired by \cite{Kwon19}, this paper aims to build such a generative model for visualizing a graph as adjacency matrices.
We identify unique challenges for designing a neural network to achieve the goal and introduce a novel encode-decoder architecture that can learn a latent space of diverse matrix reorderings.

\subsection{Deep Learning for Sorting}
\label{dmr:sec:related-work:soft-sort}
Sorting is one of the most important problems in computer science.
Matrix reordering is also a sorting problem, where the goal is to find an ordering that can uncover structural information in a relational matrix.
For our goal, we need to train a neural network that can produce diverse reorderings of the adjacency matrix of the given graph.
However, sorting is not differentiable with respect to the input.
It is thus impossible to use end-to-end gradient-based optimization to train a neural network with exact sorting functions.

Recently, a number of relaxed (differentiable) sorting operators have been introduced to enable gradient-based learning of models that include a sorting process \cite{Sinkhorn, NeuralSort, SoftSort}.
The outputs of these relaxed sorting operators are ``soft'' permutation matrices.
For example, the output of the Sinkhorn operator \cite{Sinkhorn} is a doubly stochastic matrix, where the sum of any row or any column is 1.
In addition, the NeuralSort \cite{NeuralSort} and SoftSort \cite{SoftSort} operators produce a unimodal row stochastic matrix, where any row has the sum of 1 and has a unique $\operatorname{argmax}$.

Using soft permutation matrices computed by the relaxed sorting operators, we can train a neural network model with approximated solutions.
Once training is completed, these relaxed sorting operators can output ``exact'' permutation matrices to produce actual solutions for evaluation.
For instance, since an output of the NeuralSort \cite{NeuralSort} and SoftSort \cite{SoftSort} operators is a unimodal row stochastic matrix, we can simply apply row-wise $\operatorname{argmax}$ to get an exact permutation matrix.
For a model with the Sinkhorn operator \cite{Sinkhorn}, an exact permutation can be obtained by passing the output of Sinkhorn operator to a linear assignment problem solver (e.g., Hungarian method \cite{Hungarian}).

We use these relaxed sorting operators for learning a deep generative model that produces diverse reorderings of the adjacency matrix of a graph.
In the evaluation, we compare two relaxed sorting operators for our problem, namely Sinkhorn \cite{Sinkhorn} and SoftSort \cite{SoftSort}.
\section{Approach}
\label{dmr:sec:approach}
This paper introduces a technique for users to intuitively find the desired matrix view of a graph.
More specifically, the goal is to learn a deep generative model that can systematically produce multiple adjacency matrix views of a graph with diverse node orderings that can reveal different structural information of the same graph.
This section describes the whole process of constructing a deep generative model for matrix reordering, from gathering training samples to designing the neural network architecture.

\subsection{Collecting Training Data}
\label{dmr:sec:approach:training_data}
In order to learn a high-quality generative model that can produce various matrix reorderings, we need to collect a large and diverse set of matrix reorderings of the given graph.
We collect such a dataset using various matrix reordering methods, multiple measures for computing distances between nodes, and random initial node orderings.

\begin{figure}[t]
\centering
\captionsetup{farskip=0pt}
\captionsetup[subfigure]{justification=centering,labelformat=empty}
\makeatletter
\@wholewidth0.2pt
\makeatother
\subfloat[VAT]{\frame{\includegraphics[width=0.235\linewidth,keepaspectratio]{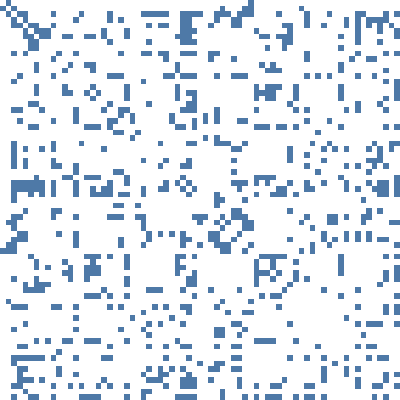}}}\hfill
\subfloat[Spectral]{\frame{\includegraphics[width=0.235\linewidth,keepaspectratio]{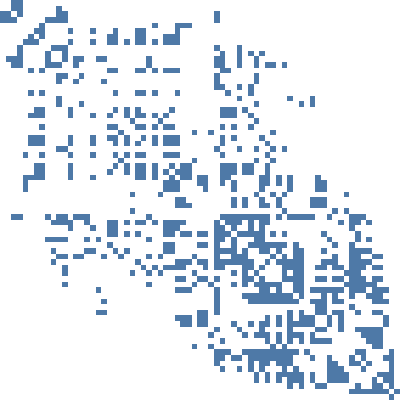}}}\hfill
\subfloat[HC-Average]{\frame{\includegraphics[width=0.235\linewidth,keepaspectratio]{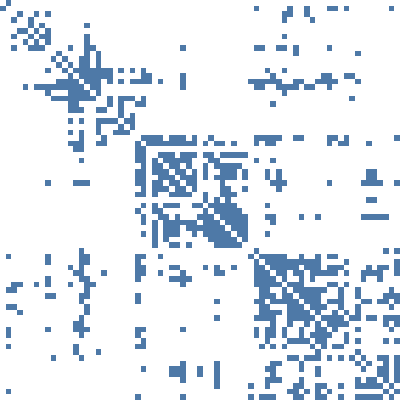}}}\hfill
\subfloat[QAP-BAR]{\frame{\includegraphics[width=0.235\linewidth,keepaspectratio]{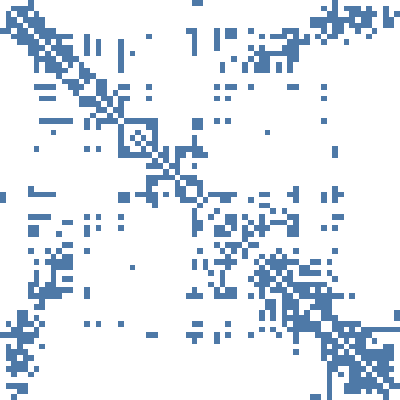}}}
\caption{Different structural patterns of the same graph can appear depending on the reordering methods.
This example shows four different reorderings of the adjacency matrix of the macaque cortical connectivity (\autoref{dmr:tab:data}) using the shortest-path distances between the nodes.
}
\label{dmr:fig:comparison-reord-methods}
\end{figure}

As we discussed in \autoref{dmr:sec:related-work:reord}, decades of research have led to many matrix reordering methods.
The adjacency matrix of a graph can reveal different structural patterns depending on which method is used, as shown in \autoref{dmr:fig:comparison-reord-methods}.
Thus, an obvious way to collect diverse reorderings of the given graph is using multiple reordering methods.

\begin{figure}[t]
\centering
\captionsetup{farskip=0pt,skip=0pt}
\includegraphics[width=\linewidth]{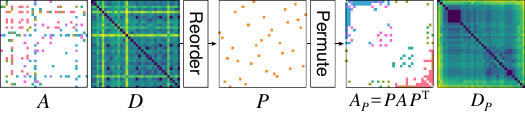}
\vspace{-1em}
\caption{Reordering the adjacency matrix $A$ of an undirected graph.
A reordering method takes pairwise dissimilarities $D$ between the nodes and computes a permutation $P$ of the nodes minimizing a loss function to reveal structural information in the graph.
In this example, two primary clusters (blue and red) are clearly visible in the reordered adjacency matrix $A_P$ but not in the unordered one $A$.}
\label{dmr:fig:seriate}
\end{figure}

This paper focuses on a simple, undirected, unlabeled graph.
Since an adjacency matrix of an undirected graph is a symmetric matrix, we can use a single node ordering for sorting both the rows and columns of the matrix.
\autoref{dmr:fig:seriate} shows an overview of the matrix reordering process.
To reorder the adjacency matrix $A$ of an undirected graph,
a matrix reordering method typically begins with a pairwise distance (or dissimilarity) matrix $D$, where $D_{i,j}$ represents the distance between nodes $u_i$ and $u_j$, and $D_{i,i} = 0$.
Thus, depending on the measure of distance, we can get different reordering results with the same reordering method.

\begin{figure}[b]
\vspace{-1em}
\captionsetup[subfigure]{justification=centering,labelformat=empty}
\centering
\makeatletter
\@wholewidth0.2pt
\makeatother
\subfloat[Shortest-Path]{\frame{\includegraphics[width=0.235\linewidth,keepaspectratio]{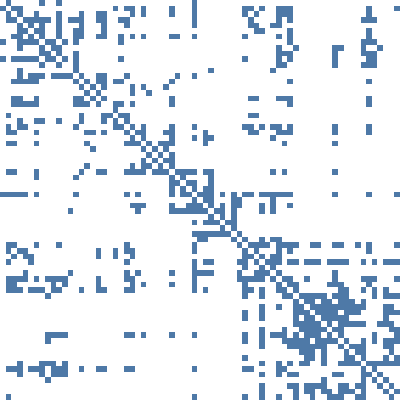}}}\hfill
\subfloat[Euclidean]{\frame{\includegraphics[width=0.235\linewidth,keepaspectratio]{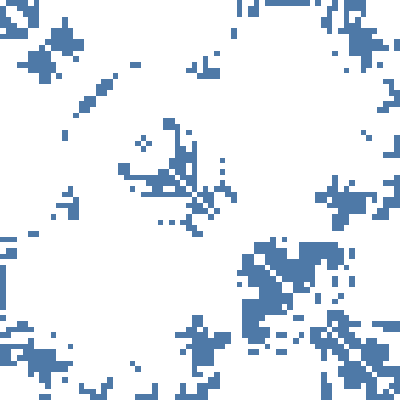}}}\hfill
\subfloat[Jaccard]{\frame{\includegraphics[width=0.235\linewidth,keepaspectratio]{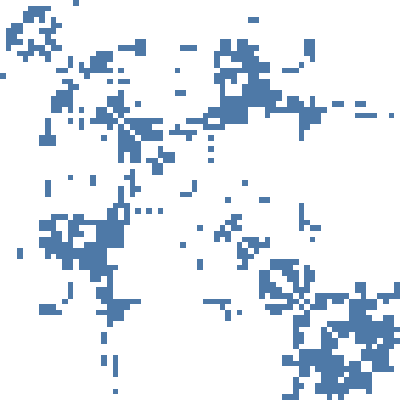}}}\hfill
\subfloat[Yule]{\frame{\includegraphics[width=0.235\linewidth,keepaspectratio]{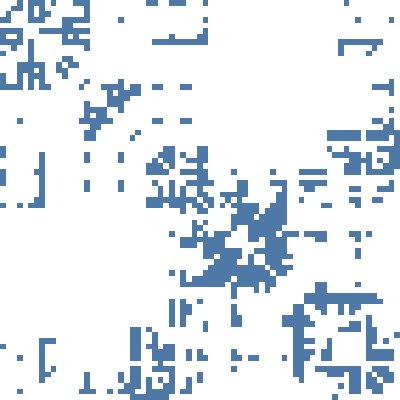}}}
\caption{Different structural patterns of the same graph can appear depending on the distance measure, although the same reordering method is used.
This example shows reordering results of the \textsf{macaque} graph (\autoref{dmr:tab:data}) using three different distance measures with the TSP \cite{TSPOrdering, Climer06} method.
}
\label{dmr:fig:comparison-dist-measures}
\end{figure}

The shortest-path distance is one of the most common measures for calculating dissimilarity between nodes in a graph.
Another common approach is considering the adjacency matrix of a graph as the feature vectors of nodes, where each row/column represents the corresponding node.
Then, general distance (or dissimilarity) functions (e.g., \autoref{dmr:tab:distances}) can be used to obtain the pairwise distance matrix, and the resulting matrix reorderings can vary greatly depending on the measure, as shown in \autoref{dmr:fig:comparison-dist-measures}.
We include the shortest-path distance and various distance functions (e.g., \autoref{dmr:tab:distances}) to collect diverse reordering results of a graph.

\begin{figure}[t]
\captionsetup{farskip=0pt}
\captionsetup[subfigure]{justification=centering,labelformat=empty}
\centering
\makeatletter
\@wholewidth0.2pt
\makeatother
\subfloat[ARSA with $A$]{\frame{\includegraphics[width=0.235\linewidth,keepaspectratio]{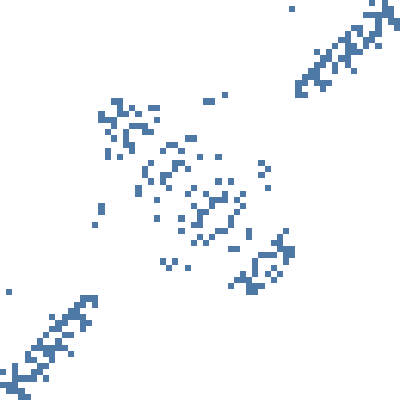}}}\hfill
\subfloat[ARSA with $\hat{A}$]{\frame{\includegraphics[width=0.235\linewidth,keepaspectratio]{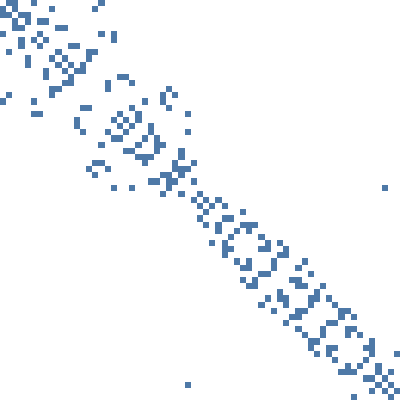}}}\hfill\hfill
\subfloat[R2E with $A$]{\frame{\includegraphics[width=0.235\linewidth,keepaspectratio]{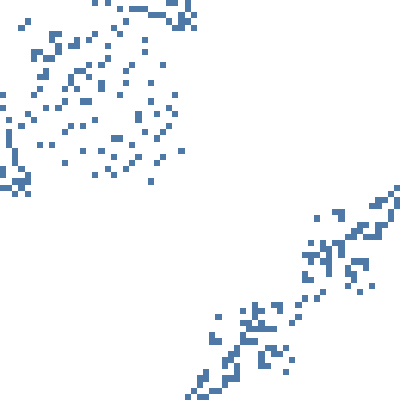}}}\hfill
\subfloat[R2E with $\hat{A}$]{\frame{\includegraphics[width=0.235\linewidth,keepaspectratio]{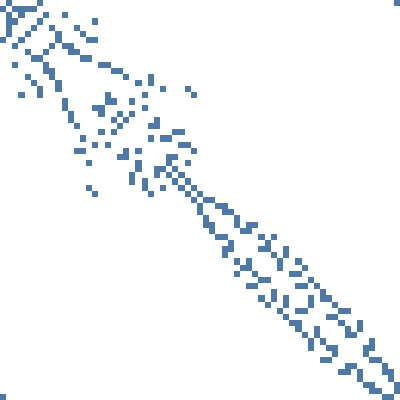}}}
\caption{Different structural patterns of the same graph can appear by adding self-loops even with the same reordering algorithm and the same distance function.
In this example, the matrix reorderings of the \textsf{gd96c} graph (\autoref{dmr:tab:data}) are computed with ARSA \cite{Brusco08} and R2E \cite{R2E} and the Jaccard distances between nodes.
The results are quite different depending on whether the raw adjacency matrix ($A$) or the adjacency matrix with self-loops ($\hat{A}$) is used for computing distances between nodes.
The adjacency matrix with self-loops ($\hat{A}$) is only used for computing distances between nodes but not for visualizing a graph.}
\label{dmr:fig:comparison-self-loops}
\end{figure}

When two nodes have the same set of adjacent nodes, their adjacency vectors (i.e., a row/column of the adjacency matrix) are the same.
In this case, the distance between the two nodes can be calculated as 0, if the distance function only considers their adjacency vectors (e.g., Euclidean distance).
In order to distinguish nodes with the same set of adjacent nodes, self-loops are often added (i.e., set the diagonal elements of the adjacency matrix to be 1) before computing the pairwise distance matrix, so that the distance between two different nodes is greater than 0.
The same reordering method with the same distance function can produce very different results depending on whether self-loops are added or not, as shown in
\autoref{dmr:fig:comparison-self-loops}.
We use both the (raw) adjacency matrix $A$ and the adjacency matrix with self-loops $\hat{A}$ of the given graph when we compute the pairwise distances.

\begin{figure}[b]
\vspace{-1em}
\captionsetup[subfigure]{justification=centering,labelformat=empty}
\centering
\makeatletter
\@wholewidth0.2pt
\makeatother
\subfloat[(a)]{\frame{\includegraphics[width=0.235\linewidth,keepaspectratio]{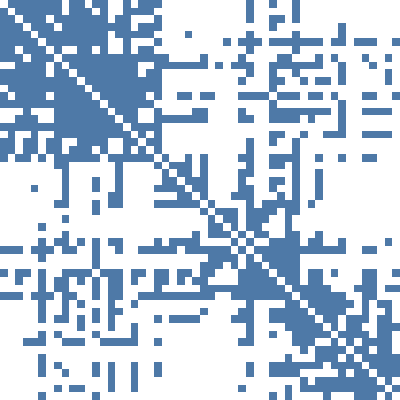}}}\hfill
\subfloat[(b)]{\frame{\includegraphics[width=0.235\linewidth,keepaspectratio]{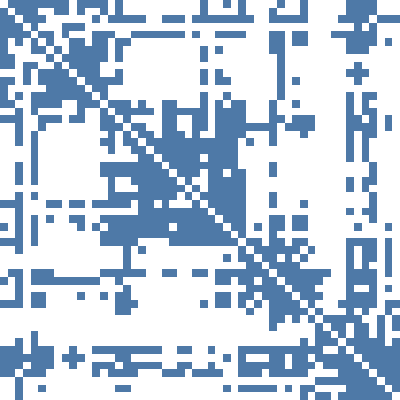}}}\hfill
\subfloat[(c)]{\frame{\includegraphics[width=0.235\linewidth,keepaspectratio]{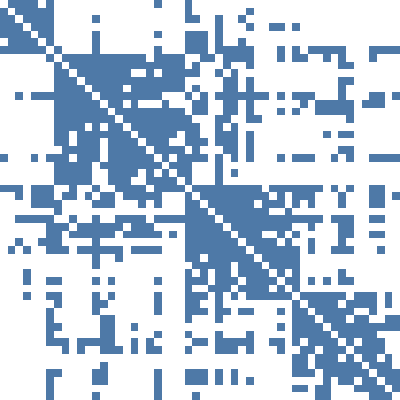}}}\hfill
\subfloat[(d)]{\frame{\includegraphics[width=0.235\linewidth,keepaspectratio]{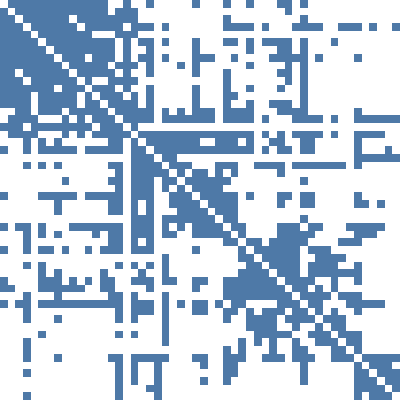}}}
\caption{Different structural patterns of the same graph can appear depending on the initial node ordering even though the same reordering method and the distance measures between nodes are used (OLO-Average and shortest-path distance in this example).
In the previous examples (\autoref{dmr:fig:comparison-reord-methods}, \autoref{dmr:fig:comparison-dist-measures}, \autoref{dmr:fig:comparison-self-loops}), the same initial node orderings were used for each graph.}
\label{dmr:fig:comparison-rand-init-ord}
\end{figure}
Lastly, several reordering methods often produce different results depending on the initial node orderings.
\autoref{dmr:fig:comparison-rand-init-ord} shows such an example, where the same reordering method with the same distance measure produces different reordering results just because of the initial orderings of the given adjacency matrix are different.
Thus, we also include a number of random initial orderings when collecting a training dataset.

Obtaining a large collection of matrix reorderings can take a significant amount of time.
However, the training process can begin as soon as we have a few dozen samples with mini-batch training schemes.
Therefore, the model training process and the data collection process can be performed simultaneously and incrementally.
With the incremental training process, users can utilize our generative model as quickly as possible.

\subsection{Permutation Equivariance}
\label{dmr:sec:approach:perm-eq}
A permutation of the nodes of a graph is said to be an automorphism if the permutation preserves the adjacency of each node.
In other words, if the permutation $P$ is an automorphism, the permuted adjacency matrix $A_P = PAP^{\textsf{T}}$ is equal to the original adjacency matrix $A$ ($0 = A_P - A$).

\begin{figure}[b]
\vspace{-1em}
\centering
\makeatletter
\@wholewidth0.2pt
\makeatother
\subfloat[$A$]{\frame{\includegraphics[width=0.235\linewidth,keepaspectratio]{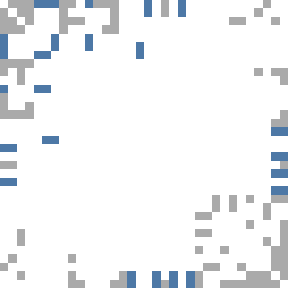}}}\hfill
\subfloat[$P_1$]{\frame{\includegraphics[width=0.235\linewidth,keepaspectratio]{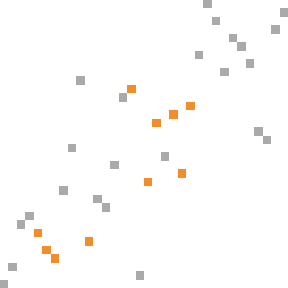}}}\hfill
\subfloat[$P_2$]{\frame{\includegraphics[width=0.235\linewidth,keepaspectratio]{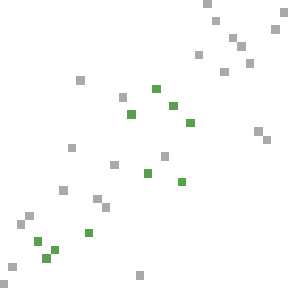}}}\hfill
\subfloat[$A_{P_1}$ and $A_{P_2}$]{\frame{\includegraphics[width=0.235\linewidth,keepaspectratio]{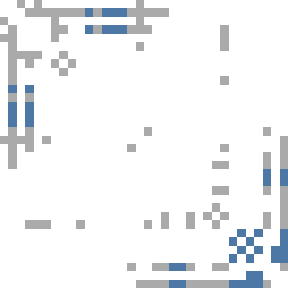}}}
\caption{Permutation equivariance.
Different permutations can produce an equal matrix reordering result depending on the structure of the given graph.
In this example, although the permutations $P_1$ and $P_2$ are different, the reordered adjacency matrices $A_{P_1}$ and $A_{P_2}$ are equal.
The orange and green cells are the difference in the two permutations $P_1$ and $P_2$ (b and c), and the blue cells (a and d) are the incident edges of the nodes in the graph associated with the difference.}
\label{dmr:fig:perm-eq}
\end{figure}

When a graph has a nontrivial automorphism (i.e., non-identity automorphisms), multiple different permutations of the graph can produce an equal reordered adjacency matrix.
\autoref{dmr:fig:perm-eq} shows such an example, where two different permutations $P_1$ and $P_2$ of the \textsf{karate} graph produce the same reordered adjacency matrix ($0 = A_{P_1} - A_{P_2}$).

Hence, we need to design a neural network architecture and a loss function considering the fact that different permutations can produce an equal reordered adjacency matrix.
However, finding all the permutations that produce the same reordered adjacency matrix is practically impossible since the graph automorphism problem belongs to the class NP of computational complexity, similar to the graph isomorphism problem.
Therefore, the architecture and the loss function need to be permutation equivariant:
they should not depend on how the nodes are permuted but only on the resulting reordered adjacency matrices.
We thus train the model to reconstruct reordered adjacency matrices, not to reconstruct permutations (i.e., node orderings).
We describe our design of the architecture and the loss function in more detail in \autoref{dmr:sec:approach:architecture}.

\subsection{Architecture}
\label{dmr:sec:approach:architecture}
\begin{figure*}[t]
\captionsetup{farskip=0pt,skip=0pt}
\centering
\includegraphics[width=\textwidth, keepaspectratio]{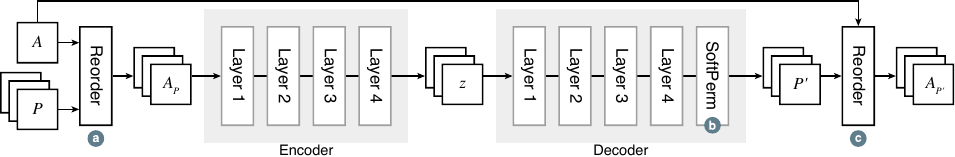}
\caption{%
Our encoder-decoder architecture that learns a generative model from a collection of example matrix reorderings.
We describe this architecture in \autoref{dmr:sec:approach:architecture}}
\label{dmr:fig:architecture}
\end{figure*}

This section discusses an encoder-decoder architecture to learn a generative model that can produce diverse matrix reorderings of the given graph.
The architecture follows the general framework of VAEs \cite{VAE}.
\autoref{dmr:fig:architecture} shows an overview of the architecture.

\subsubsection{Encoder}
The encoder's goal is to learn the low-dimensional latent representation $z$ of a reordered adjacency matrix $A_P = \operatorname{Reorder}(A, P) = P A P^{\textsf{T}}$ (\autoref{dmr:fig:architecture}a), where $A$ is the original adjacency matrix and $P$ is the permutation matrix that represents a node ordering.
As we discussed in \autoref{dmr:sec:approach:perm-eq}, we use a reordered adjacency matrix $A_P$ as the input object rather than a permutation $P$ because there can be different permutations that produce the same reordered adjacency matrix.

There is another intuition of using reordered adjacency matrices as the input objects:
we can treat a reordered adjacency matrix as an image since it is a visual representation of the given graph designed to help the viewers comprehend the data.
With this intuition, we design the encoder inspired by neural networks for learning a generative model of images \cite{VAE, DCGAN}.
For example, Kingma and Welling \cite{VAE} have designed a variational autoencoder using a multi-layer perceptron (a fully-connected neural network), where the input images are converted as one-dimensional vectors before they are fed to the encoder.
In addition, Radford et al. \cite{DCGAN} used a convolutional neural network, which has been widely used for various computer vision tasks \cite{DeepLearningNature}, for designing a generative adversarial network \cite{GAN}.

Whether we use a multi-layer perceptron (MLP) or a convolutional neural network (Conv), the fundamental goal of the encoder is to learn a highly compressed representation $z$ that captures the essence of a reordered adjacency matrix $A_P$, where each layer of the encoder gradually reduces the dimensionality of representations.
In the evaluation (\autoref{dmr:sec:eval}), we compare two encoder designs (MLP and Conv) for learning a generative model of matrix reordering.

We set the encoder to produce a two-dimensional representation of the input $A_P$.
Similar to the latent space of node-link diagrams \cite{Kwon19}, a two-dimensional latent space is straightforward to map a grid of generated samples on the latent space.
Thus, the users can interpret and navigate with ease the latent space by using the grid of generated samples as a WYSIWYG interface.
We use the uniform distribution of $[-1, 1]^2$ as the prior distribution so that the encoder learns to represent the input $A_P$ in $[-1, 1]^2$.

\subsubsection{Decoder}
The decoder's goal is to reconstruct the input objects from their latent representations---computed by the encoder.
In general, the decoder is trained to reconstruct the input objects (e.g., images of handwritten digits \cite{VAE}) directly from the trainable parameters (i.e., weights and biases) of the neural network.
However, for an adjacency matrix, naively training the decoder to reconstruct an input (i.e., a reordered adjacency matrix $A_P$) from the decoder leads to producing an invalid adjacency matrix.

For general image data (e.g., landscapes, portraits, and animals), small errors in the generated results do not have a substantial effect on the viewers' understanding of the image content.
In the case of an adjacency matrix view, however, even a small reconstruction error can lead to producing an invalid visualization that can mislead the viewers about the data, where the graph structure represented by a reconstructed adjacency matrix is different from the structure of the given graph.
Therefore, we need to ensure that the decoder produces adjacency matrices representing exactly the same structure as the given graph.

We design the decoder to generate a permutation matrix $P'$ and then reorder the given adjacency matrix $A$ with the generated permutation matrix $P'$.
Specifically, the input $A_P$ is reconstructed as $A_{P'} = \operatorname{Reorder}(A, P') = P'\! A P'^{\textsf{T}}$, where $P'$ is the output of the decoder and $A$ is the original adjacency matrix (\autoref{dmr:fig:architecture}c).
In essence, the decoder learns to reconstruct the permutation matrix $P$ (or its automorphism) of the input matrix reordering $A_P = P A P^{\textsf{T}}$, not to directly reconstruct $A_P$.

For the decoder to produce a permutation matrix, we use a relaxed sorting operator that outputs permutation matrices, such as Sinkhorn~\cite{Sinkhorn} and SoftSort~\cite{SoftSort} (discussed in \autoref{dmr:sec:related-work:soft-sort}).
A relaxed sorting operator is placed after the decoder's layers with trainable parameters (SoftPerm \autoref{dmr:fig:architecture}b).
We set the SoftPerm to produce ``soft'' permutation matrices during training to enable gradient-based learning and exact permutation matrices at the inference time.
While a reconstructed adjacency matrix $A_{P'}$ can be different from the input $A_P$, the decoder always produces a valid permutation matrix.
Therefore, the decoder is guaranteed to generate adjacency matrices that represent the same structure as the given graph.

While both Sinkhorn~\cite{Sinkhorn} and SoftSort~\cite{SoftSort} operators produce a permutation matrix, they have several differences in terms of their computational complexity and the mathematical characteristic of produced ``soft'' permutation matrices.
The Sinkhorn operator $\mathcal{S}(X)$ over an $n \times n$ square matrix $X$ is defined as:
\begin{eqnarray}
\nonumber \mathcal{S}^0(X) &=& \exp(X), \\
\nonumber \mathcal{S}^l(X) &=& \mathcal{T}_c\left(\mathcal{T}_r\left(\mathcal{S}^{l-1}\left(X\right)\right)\right),\\
\nonumber \mathcal{S}(X) &=&\lim_{l\rightarrow \infty} \mathcal{S}^l(X). 
\end{eqnarray}
where $\mathcal{T}_r(X)$ and $\mathcal{T}_c(X)$ are the row- and column-wise normalization operators, respectively.
On the other hand, the SoftSort operator takes an $n$-dimensional vector $h$ and produces a permutation matrix by:
\begin{equation*}
    \operatorname{softmax}\left(\frac{-d\left(\operatorname{sort}(h) \mathds{1}^\mathsf{T}, \mathds{1} h^\mathsf{T}\right)}{\tau}\right)
\end{equation*}
where $\operatorname{softmax}$ is applied row-wise, $d$ is a differentiable semi-metric function, and $\tau$ is a temperature parameter for controlling the degree of relaxation.
In addition, to get an exact permutation matrix at the inference stage, the Sinkhorn operator requires to solve a linear assignment problem. Contrastingly, the SoftSort operator only needs to apply $\operatorname{argmax}$ row-wise.
Therefore, the Sinkhorn operator is more computationally demanding than the SoftSort operator.
We compare two decoder designs in the evaluation to investigate their differences in building a generative model of matrix reordering.

\subsection{Training}
The parameters of the encoder and the decoder are trained to minimize the reconstruction loss $L_X$ and the variational loss $L_Z$, as our architecture is a variational autoencoder \cite{VAE}.
For the variational loss $L_Z$, we use the sliced-Wasserstein distance \cite{SWAE}, like the generative model for node-link diagrams \cite{Kwon19}.

The reconstruction loss $L_X$ quantifies the difference between the input matrix reordering $A_P = P A P^{\mathsf{T}}$ and its reconstruction $A_{P'} = P'\! A P'^{\mathsf{T}}$, where $A$ is the original adjacency matrix, $P$ is the permutation matrix that represents the input node ordering, and $P'$ is the permutation matrix produced by the decoder.
We don't consider the difference between the input node ordering $P$ and the generated node permutation $P'$ since there can be different orderings that produce the same matrix reordering of the input adjacency matrix, as we discussed in \autoref{dmr:sec:approach:perm-eq}.
We directly compare the difference between the corresponding elements of $A_P$ and $A_{P'}$.
For example, we can use the sum of the differences as the reconstruction loss: $L_X = \sum\limits_{i,j} \|(A_P)_{i,j} - (A_{P'})_{i,j} \|$.
In addition, since the adjacency matrix of a simple graph is a binary matrix, the binary cross-entropy loss can be used: $L_X = \sum\limits_{i,j}\ -\Big((A_P)_{i,j} \cdot \log (A_{P'})_{i,j} + \left(1 - (A_P)_{i,j}\right) \cdot \log \left(1 - \log (A_{P'})_{i,j}\right)\Big)$.

\section{Evaluation}
\label{dmr:sec:eval}
The primary goal of the evaluation is to validate whether the model can learn to produce diverse matrix reorderings of the given graphs, not just memorize the training samples. 
We perform quantitative and qualitative evaluations of the generative model for reconstructing unseen matrix reorderings (i.e., the test dataset).
\subsection{Datasets}
\label{dmr:sec:eval:data}
This evaluation uses eight real-world graphs and around ten thousand orderings per graph (\autoref{dmr:tab:data}).
While the number of training samples could affect the model's behaviors, this evaluation focuses on the effect of different architectures using a fixed number of training samples per graph.
For each graph, we first collected 13,500 matrix reorderings using
25 distance measures between nodes (\autoref{dmr:tab:distances}),
27 matrix reordering methods (\autoref{dmr:tab:reordering_methods}),
and 20 random initial orderings.

\begin{table}[t]
\caption{The eight graphs used in the evaluation.
$|V|$:~the number of nodes,
$|E|$:~the number of edges,
$|A_P|$:~the number of unique reordering (Section \ref{dmr:sec:eval:data}),
$t$:~the time it took to compute the training dataset in minutes}
\label{dmr:tab:data}
{\setstretch{1.1}
\setlength{\tabcolsep}{0.575em}
\centering
\footnotesize{
\begin{tabu}{@{}l l r r r r r r@{}}
Name                  & Type             & $|V|$ & $|E|$ & $|A_P|$ & $t$   & Source\\\hline
\textsf{karate}       & Social           & 34    & 78    & 7,964   & 3.27  & \cite{konect, Karate}\\
\textsf{cat}          & Brain            & 52    & 515   & 9,060   & 4.49  & \cite{CatCortex}\\
\textsf{gd96c}        & Synthetic        & 65    & 125   & 10,973  & 6.65  & \cite{GD96C}\\
\textsf{macaque}      & Brain            & 71    & 438   & 9,682  & 7.16  & \cite{Young93}\\
\textsf{polbooks}     & Co-purchase      & 105   & 441   & 10,303  & 11.15 & \cite{konect}\\
\textsf{collab}       & Collaboration    & 129   & 847   & 8,420  & 13.68 & \cite{Yanardag15}\\
\textsf{ego-fb}       & Social           & 148   & 1,692 & 10,193  & 16.15 & \cite{SNAP, EgoFacebook}\\
\textsf{jazz}         & Collaboration    & 198   & 2,742 & 10,212  & 27.58 & \cite{konect, Jazz}\\
\end{tabu}
}
}
\end{table}

The 25 distance measures consist of the shortest-path distances between nodes, 12 distance measures (\autoref{dmr:tab:distances}) based on the adjacency matrix (a row of adjacency matrix represents a node), and the 12 distance measures based on the adjacency matrix with self-loops (i.e., the diagonal elements of the matrix is 1).
The 27 matrix reordering methods (\autoref{dmr:tab:reordering_methods}) are selected from the R package \textsf{seriation} \cite{RSeriation}, which provides publicly available, robust implementations of multiple matrix reordering methods.
Although other methods are available in the package, these methods could not compute reorderings of all the eight graphs in a reasonable amount of time (24 hours).
The 20 random initial node orderings are used as some matrix reordering methods produce different results depending on the initial node ordering of the given graph (\autoref{dmr:fig:comparison-rand-init-ord}).

A reordering can be the same as the reverse of another. Therefore, a reordering is reversed if its reverse has a higher similarity with a reference reordering---a Spectral reordering is used in this evaluation.

To properly evaluate the generalization capability of a model, the reorderings in the test set should not be used for training the model.
Thus, we extract unique reordered adjacency matrices from the 13,500 reorderings of a graph since there can be equal reorderings, as we discussed in \autoref{dmr:sec:approach:perm-eq}
The numbers of unique matrix reorderings and the time it took to collect the training dataset of each graph are shown in \autoref{dmr:tab:data}.
For the reconstruction evaluation, we perform 5-fold cross-validations, where 80\% of unique reorderings of a graph are used for training, and the other 20\% reorderings are used for testing.
Extracting unique adjacency matrices is required only for the evaluation purpose; it is optional for using our technique in practice.

\begin{table}[t]
\caption{The 12 metrics used in the evaluation for computing the dissimilarity between a pair of nodes. $N_{ij}$ is the number of occurrences of $u_k = i$ and $v_k = j$, and $n$ is the number of nodes. In the evaluation, 25 distance measures are derived using these metrics consisting of 12 distance measures based on the original adjacency matrix, 12 distance measures based on the adjacency matrix with self-loops (i.e., the diagonal elements of the matrix is 1), and the shortest-path distances between nodes.
}
\label{dmr:tab:distances}
\centering
{\setstretch{1.2}
\tabulinesep=0.175em
\footnotesize{
\begin{tabu}{@{}X[6,l,m] X[17,l,m]@{}}
Distance & Definition \\\hline
Euclidean & $\displaystyle \sqrt{\sum \left(u_i - v_i\right)^2} $\\
Manhattan & $\displaystyle \sum \left\vert u_i - v_i\right\vert $\\
Cosine & $\displaystyle 1 - \left( u\,v \big/ \lVert u \rVert_2 \lVert v \rVert_2 \right)$\\
Dice & $\displaystyle \left(N_{01} + N_{10}\right) \big/ \left(2N_{11} + N_{01} + N_{10}\right)$\\
Hamming & $\displaystyle \left(N_{01} + N_{10}\right) \big/ n$\\
Jaccard & $\displaystyle \left(N_{01} + N_{10}\right) \big/ \left(N_{11} + N_{01} + N_{10}\right)$\\
Kulsinski & $\displaystyle \left(N_{01} + N_{10} - N_{11} + n\right) \big/ \left(N_{01} + N_{10} + n\right)$\\
Rogers-Tanimoto & $\displaystyle 2\left(N_{01} + N_{10}\right) \big/ \big(N_{00} + N_{11} + 2\left(N_{01} + N_{10}\right)\big)$\\
Russell-Rao & $\displaystyle \left(n - N_{11}\right) \big/ n$\\
Sokal-Michener & $\displaystyle 2\left(N_{01} + N_{10}\right) \big/ \big(2\left(N_{00} + N_{11}\right)+ 2\left(N_{01} + N_{10}\right)\big)$\\
Sokal-Sneath & $\displaystyle 2\left(N_{01} + N_{10}\right) \big/ \big(N_{11} + 2\left(N_{01} + N_{10}\right)\big)$\\
Yule & $\displaystyle 2 N_{01} N_{10} \big/ \left(N_{00} N_{11} + N_{01} N_{10}\right)$
\end{tabu}
}
}
\end{table}

\begin{table}[h]
\caption{The 27 matrix reordering methods used for producing training data in the experiment. The method names follow the documentation of the R package \textsf{seration} \cite{RSeriation}.}
\label{dmr:tab:reordering_methods}
\centering
{\setstretch{1.2}
\tabulinesep=0.35em
\footnotesize{
\begin{tabu}{@{}X@{}}
\textbf{ARSA}: Minimize the linear seriation using simulated annealing \cite{Brusco08}.\\\hline
\textbf{TSP}: Minimize the Hamiltonian path length using a traveling salesperson problem solver \cite{TSPOrdering, Climer06}.\\\hline
\textbf{R2E}: Rank-two ellipse seriation method \cite{R2E}\\\hline
\textbf{MDS-$\{$Metric, Nonmetric, Angle$\}$}: Minimize stress of the linear order using multidimensional scaling (MDS) \cite{MDSKendall}. Three variants are used.\\\hline
\textbf{HC-$\{$Average, Single, Complete, Ward$\}$}: Uses a linear order of the leaf nodes in a dendrogram obtained by hierarchical clustering \cite{Eisen98}. Each variant uses a different clustering method.\\\hline
\textbf{GW-$\{$Average, Single, Complete, Ward$\}$}: Hierarchical clustering methods with the leaf-node ordering method developed by Gruvaeus and Wainer \cite{Gruvaeus72}.\\\hline
\textbf{OLO-$\{$Average, Single, Complete, Ward$\}$}: Hierarchical clustering methods with the optimal leaf ordering method by Bar-Joseph~et~al.~\cite{OLO}\\\hline
\textbf{VAT}: Uses the order of node addition in Prim's algorithm, which finds a minimum spanning tree \cite{VAT}.\\\hline
\textbf{Spectral-$\{\varnothing$, Norm$\}$}: Uses the order of the Fiedler vector \cite{SpectralOrdering, DingHe04}. ``Spectral'' uses the (unnormalized) Laplacian matrix, and ``Spectral-Norm'' uses the normalized Laplacian matrix.\\\hline
\textbf{SPIN-$\{$NH, STS$\}$}: Sorting Points Into Neighborhoods \cite{Tsafrir05}. Two variants are used: ``SPIN-NH'' uses the neighborhood algorithm, and ``SPIN-STS'' implements the side-to-side algorithm \cite{Tsafrir05}.\\\hline
\textbf{QAP-$\{$LS, 2Sum, BAR, Inertia$\}$}: Formulates reordering as a quadratic assignment problem. Each variant uses different heuristics,
namely the linear seriation problem formulation (QAP-LS) \cite{HubertSchultz76},
the 2-sum problem formulation (QAP-2Sum) \cite{SpectralOrdering},
the banded anti-Robinson form (QAP-BAR) \cite{BandedAntiRobinson},
and the inertia criterion (QAP-Inertia) \cite{PermutMatrix}.
\end{tabu}
}
}
\end{table}

\subsection{Architectures and Configurations}
\label{dmr:sec:eval:model}
\begin{figure*}[t]
\captionsetup{farskip=0pt,skip=0pt}
\centering
\includegraphics[width=\textwidth, keepaspectratio]{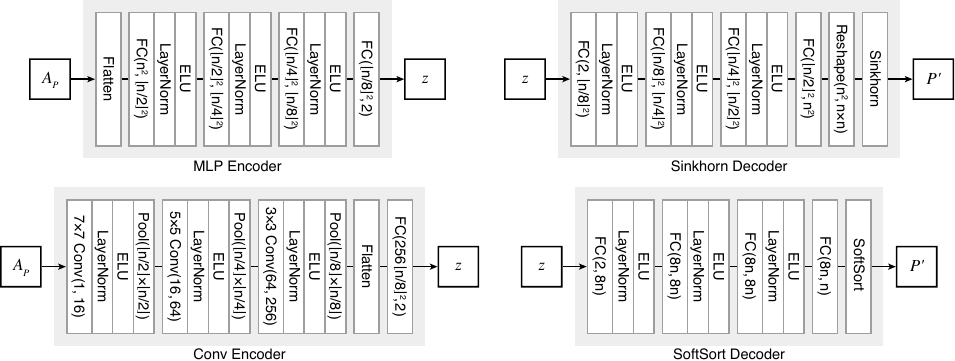}
\vspace{-1em}
\caption{%
The encoder and decoder architectures used in the evaluation. Four architectures are derived based on the combinations of these encoder and decoder designs: MLP-Sinkhorn, MLP-SoftSort, Conv-Sinkhorn, and Conv-SoftSort. \textsf{FC(m,~n)} is a fully-connected layer with the input feature size of $m$ and the output feature size of $n$. \textsf{k$\times$k~Conv(m, n)} is a convolution layer with the kernel size of $k \times k$, the input feature size of $m$, and the output feature size of $n$. \textsf{Pool(s)} is an adaptive max pooling layer with the target size of $s$. \textsf{LayerNorm} is the layer normalization \cite{LayerNorm}. \textsf{ELU} is the exponential linear unit \cite{ELU}.
}
\label{dmr:fig:experiment:architectures}
\end{figure*}

We compare four different architectures, which are the combinations of two encoder designs (MLP and Conv) and two decoder designs (Sinkhorn and SoftSort): MLP-Sinkhorn, MLP-SoftSort, Conv-Sinkhorn, and Conv-SoftSort.
\autoref{dmr:fig:experiment:architectures} shows the encoders and decoders in detail.
The encoder architectures generally follow the ones for images.
The MLP encoders' design follows the encoder of a variational autoencoder \cite{VAE}.
Conv encoders' design is similar to the discriminator of a deep convolutional generative adversarial network \cite{DCGAN}.
Both decoder architectures produce a permutation matrix but with different relaxed sorting operators (Sinkhorn~\cite{Sinkhorn} and SoftSort~\cite{SoftSort}).
In this evaluation, we selected the SoftSort~\cite{SoftSort} operator over the NeuralSort operator~\cite{NeuralSort} as they both produce a unimodal row stochastic matrix while the SoftSort operator is more computationally efficient~\cite{SoftSort}.

\textbf{MLP} encoders use four fully-connected layers, where it takes the input reordering $A_P$---an $n \times n$ matrix---as an $n^2$-dimensional vector, where $n$ is the number of nodes of the given graph.
Each layer learns a more compressed representation of the input $A_P$ by gradually reduces the dimensionality of representations.
From the first to third layers, the $i$-th fully-connected layer (\textsf{FC} in \autoref{dmr:fig:experiment:architectures}) takes an $\lfloor n/2^{(i-1)} \rfloor^2$-dimensional vector and outputs an $\lfloor n/2^i \rfloor^2$-dimensional vector representation of the input $A_P$.
For example, the first layer takes the $n^2$-dimensional representation of the input $A_P$ and learns an $\lfloor n/2 \rfloor^2$-dimensional representation.
The last layer produces the two-dimensional latent representation $z$ of the input $A_P$ from the $\lfloor n/8 \rfloor^2$-dimensional representation of it.

\textbf{Conv} encoders use three convolution layers and one fully-connected layer, where the input reordering $A_P$ is considered as an $n \times n \times 1$ tensor.
From the first to third layers, the $i$-th convolution layer (\textsf{Conv} in \autoref{dmr:fig:experiment:architectures}) has the kernel size of $\left(7-2(i-1)\right) \times \left(7-2(i-1)\right)$ and the output feature size of $16 \cdot 4^{(i-1)}$.
For example, the second layer has the kernel size of 5 and the output feature size of 64.
Each convolution layer is followed by layer normalization \cite{LayerNorm}, ELU \cite{ELU}, and the adaptive max pooling.
The $i$-th adaptive max pooling (\textsf{Pool} in \autoref{dmr:fig:experiment:architectures}) has the target size of $\lfloor n / 2^i \rfloor \times \lfloor n / 2^i \rfloor$.
Thus, the first convolution layer takes the input $A_P$ as an $n \times n \times 1$ tensor and outputs an $\lfloor n / 2^i \rfloor \times \lfloor n / 2^i \rfloor \times 16$ tensor.
The last layer computes the two-dimensional latent representation $z$ from the output of the third convolution layer by reshaping the $\lfloor n / 8 \rfloor \times \lfloor n / 8 \rfloor \times 256$ tensor as a $256\lfloor n / 8 \rfloor^2$-dimensional vector.

\textbf{Sinkhorn} decoders use four fully-connected layers and the Sinkhorn operator \cite{Sinkhorn} for computing a permutation matrix.
The fully-connected layers of this decoder design is similar to the reverse of the MLP encoder.
The first fully-connected layer takes the two-dimensional latent representation $z$ and outputs an $\lfloor n / 8 \rfloor^2$-dimensional vector.
From the second to fourth layers, the $i$-th fully-connected layer takes an $\lfloor n/2^{(5-i)} \rfloor^2$-dimensional vector and outputs an $\lfloor n/2^{(4-i)} \rfloor^2$-dimensional vector.
For example, the fourth layer takes an $\lfloor n/2 \rfloor^2$-dimensional vector and outputs an $n^2$-dimensional vector.
The output of the fourth layer is reshaped as an $n \times n$ matrix and fed to the Sinkhorn operator, which produces a ``soft'' permutation matrix.
For computing an exact permutation matrix after training, we apply the Hungarian method \cite{Hungarian} on the output of the Sinkhorn operator.

\textbf{SoftSort} decoders use four fully-connected layers and the SoftSort operator \cite{SoftSort}.
While the input of the Sinkhorn operator is an $n \times n$ matrix, the input of the SoftSort operator is an $n$-dimensional vector.
Thus, the SoftSort decoders can produce a permutation matrix with a significantly fewer number of parameters.
In this evaluation, we designed the SoftSort decoders to have $8n$-dimensional hidden representations (see \autoref{dmr:fig:experiment:architectures} for details).
The fourth layer takes an $8n$-dimensional vector and outputs an $n$-dimensional vector, which is passed to the SoftSort operator for computing a permutation matrix.
After training, an exact permutation matrix is obtained by applying the row-wise $\operatorname{argmax}$ operation to the output of the SoftSort operator.

All the architectures use layer normalization \cite{LayerNorm} and the exponential linear unit (ELU) \cite{ELU} on every hidden layer.
Batch normalization \cite{BatchNorm} is commonly used in the neural network architectures for computer vision tasks, especially with convolutional neural networks.
However, our pilot study showed that the models with batch normalization often result in numerical instability.
The elements of an adjacency matrix are binary, whereas the pixels of a natural image (e.g., landscape, portrait, or an animal) are often 24-bit RGB colors or 8-bit grayscale.
Therefore, in our case, batch normalization can lead to numerical instability due to the lack of diverse values of a feature across the instances of a batch.

We use the Adamax optimizer \cite{Adam} with a learning rate of 0.001 for all the models.
In our pilot study, Adamax---a variant of Adam based on infinity norm---showed more stable training than the standard Adam with the $L^2$ norm.
For the reconstruction loss $L_X$, we use the binary cross-entropy loss averaged over the elements of the adjacency matrix.
We set $\tau$ to 1.0 for both the Sinkhorn and SoftSort operators.
We train each model for 500 epochs.

\subsection{Implementation}
We used the R package \textsf{seration} \cite{RSeriation} to compute reorderings for the datasets (\autoref{dmr:sec:eval:data}).
The models are implemented with PyTorch\cite{PyTorch}.
To generate the datasets and run the evaluation, we used a machine with an Intel i7-5960X (8 cores at 3.0 GHz) CPU and an NVIDIA Titan RTX GPU.
\subsection{Quantitative Evaluation}
\label{dmr:sec:eval:quantitative}
To quantitatively evaluate the generalization capabilities, we compare four different architectures (\autoref{dmr:sec:eval:model}) for each graph in terms of the average reconstruction error rate of the test dataset---around two thousand reorderings that are not used for training the models.
Specifically, a trained model takes an input matrix reordering $A_P$ that is not used for training the model, encodes it to the latent space, and then reconstructs a matrix reordering $A_{P'}$ from the latent representation.
The difference between an input matrix reordering $A_P$ and its reconstruction $A_{P'}$ of the test dataset quantifies the generalization capability of a generative model as it estimates how close the model can produce unseen matrix reorderings.
For a single reconstruction, the error is measured by the ratio of the different elements of the input matrix $A_P$ and its reconstruction~$A_{P'}$: $ \Big(\sum\limits_{i,j} \mathbbm{1}_{(A_P)_{i,j} \neq (A_{P'})_{i,j}}\Big) / n^2$.
We repeat 5-fold cross-validations ten times and report the mean error rates to reduce the effects of fold assignments.
We also compare the four architectures' computational costs based on the training time and the peak GPU memory usage, which measures the GPU memory capacity required for training.

\begin{figure*}[t]
\captionsetup{captionskip=4pt}
\centering
\subfloat[Error rate]{\includegraphics[width=0.9\textwidth, keepaspectratio]{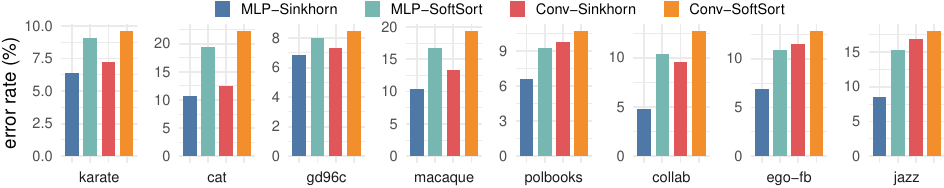}}\\
\subfloat[Training time per epoch]{\includegraphics[width=0.9\textwidth, keepaspectratio]{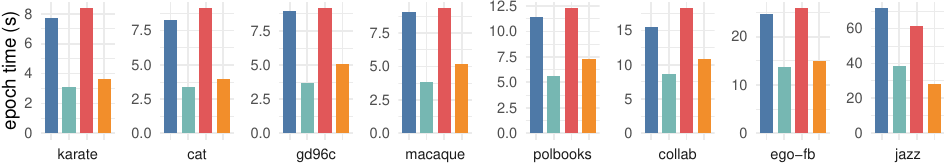}}\\
\subfloat[Peak GPU memory usage]{\includegraphics[width=0.9\textwidth, keepaspectratio]{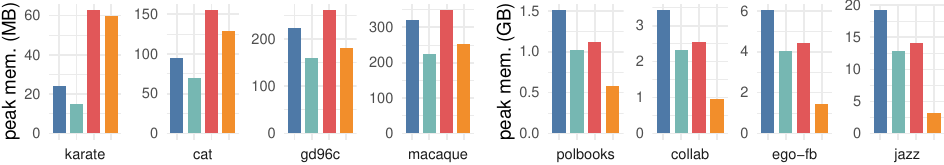}}
\caption{%
The quantitative evaluation results comparing the four architectures for each dataset.
The error rates are computed by the ratio of the difference between the input reordering $A_P$ and the reconstructed reordering $A_{P'}$.
The standard deviations (or standard errors) are not shown as they are negligible.}
\label{dmr:fig:quantitative}
\end{figure*}

\subsubsection{Results}
\autoref{dmr:fig:quantitative}a shows the mean reconstruction error rates of the 10 trials of the 5-fold cross-validations.
Overall, MLP-Sinkhorn shows the lowest error rates for all eight graphs.
The results show two different rankings of the other three architectures, depending on the graph.
For \textsf{polbooks}, \textsf{ego-fb}, and \textsf{jazz},
the models with MLP encoders show lower error rates than the ones using Conv encoders, where the ranking is as follows (from lowest to highest): MLP-Sinkhorn, Conv-Sinkhorn, MLP-SoftSort, and Conv-SoftSort.
For the other graphs, the models using Sinkhorn decoders show lower error rates than the ones with SoftSort decoders: MLP-Sinkhorn, Conv-Sinkhorn, MLP-SoftSort, and Conv-SoftSort.

\autoref{dmr:fig:quantitative}b shows the mean training time per epoch.
For all graphs, training the models with SoftSort decoders took a significantly shorter time than those with Sinkhorn decoders.
Given the same encoder, training the SoftSort decoders took about half the time it took to train the Sinkhorn decoders, on average, 50.2\% shorter.
Except for the largest graph (\textsf{jazz}), the MLP encoders took, on average, 13.3\% shorter to train than the Conv encoders with the same decoder.
Thus, for these graphs, the ranking in terms of the training time is as follows (from shortest to longest): MLP-SoftSort, Conv-SoftSort, MLP-Sinkhorn, and Conv-Sinkhorn.
For \textsf{jazz}, training the MLP encoders took, on average, 26.7\% longer than the Conv encoders given the same decoder: Conv-SoftSort (28.4~s), MLP-SoftSort (38.8~s), Conv-Sinkhorn (61.6~s), and MLP-Sinkhorn (71.8~s).

\autoref{dmr:fig:quantitative}c shows the peak GPU memory usage for training each model, which measures the required GPU memory capacity.
Given the same encoder, the models using SoftSort decoders use, on average, 36.9\% less GPU memory than those using Sinkhorn decoders.
For the graphs with less than 100 nodes (\textsf{karate}, \textsf{cat}, \textsf{gd96c}, and \textsf{macaque}), the MLP encoders use less GPU memory than the Conv encoders with the same decoder.
The difference decreases as the number of nodes increases; it becomes the opposite for the other graphs with more than 100 nodes (\textsf{polbooks}, \textsf{collab}, \textsf{ego-fb}, and \textsf{jazz}), where the Conv encoders use less GPU memory than the MLP encoders given the same decoder.

\subsubsection{Discussion}
For many computer vision tasks, including image generation \cite{DCGAN, StyleGAN2}, convolutional neural networks (CNNs) generally outperform fully-connected neural networks (i.e., multilayer perceptrons), even though a CNN often has a significantly fewer number of trainable parameters.
However, in this evaluation, the MLP encoders consistently show lower error rates than the Conv encoders when using the same decoder.
A further study using explainable artificial intelligence techniques \cite{VAinDL} is required to better understand the main cause of the accuracy differences.
However, this difference suggests that matrix reordering requires capturing global patterns of the \emph{images} (i.e., reordered adjacency matrices), which is what CNNs are not good at as they process a local neighborhood at a time \cite{NonLocal, SAGAN}.

While the MLP encoders are able to capture global patterns, they are not scalable in terms of the number of nodes, as shown in the GPU memory usage (\autoref{dmr:fig:quantitative}c).
Thus, a memory-efficient architecture is necessary to support larger graphs.
For images, several non-local modules have been introduced to understand global image structures by capturing long-range dependencies between pixels \cite{NonLocal, SAGAN, StdAlnSelfAttn}.
Although general non-local modules also consume a significant amount of memory, there have been some attempts to simplify non-local modules for images \cite{GCNet, EffAttn} to reduce memory usage.
Moreover, we believe taking the sparsity of real-world networks (i.e., an adjacency matrix is often a sparse matrix) \cite{Barabasi16} into account could drastically decrease memory consumption.

The Sinkhorn decoders show lower error rates than the SoftSort decoders given the same encoder.
This could be simply because the Sinkhorn decoders have more trainable parameters than the SoftSort decoders.
However, the experiments of Grover et al. \cite{NeuralSort} have shown that the NeuralSort operator \cite{NeuralSort}---which shares many characteristics with the SoftSort operator \cite{SoftSort}---outperforms the Sinkhorn operator for sorting tasks, although the models using the NeuralSort operator have a significantly smaller number of parameters than those with the Sinkhorn operator.
While both the Sinkhorn and SoftSort operators produce permutation matrices, the Sinkhorn operator is designed for a general matching problem (e.g., align, canonicalize, and sort), whereas the SoftSort (and NeuralSort) operator is designed specifically for sorting the input values (i.e., $\operatorname{argsort}$).
Our encoder-decoder architecture and loss function are designed more like a matching problem (e.g., jigsaw puzzles \cite{Sinkhorn}), where the goal is to produce a permutation matrix that \emph{matches} the input matrix reordering.
This design difference can affect the flow of gradients and thus causes the performance differences.
\subsection{Qualitative Results}
\label{dmr:sec:eval:qualitative}

\begin{figure*}[t]
\captionsetup{farskip=0pt,skip=0pt}
\centering
\includegraphics[width=\textwidth, keepaspectratio]{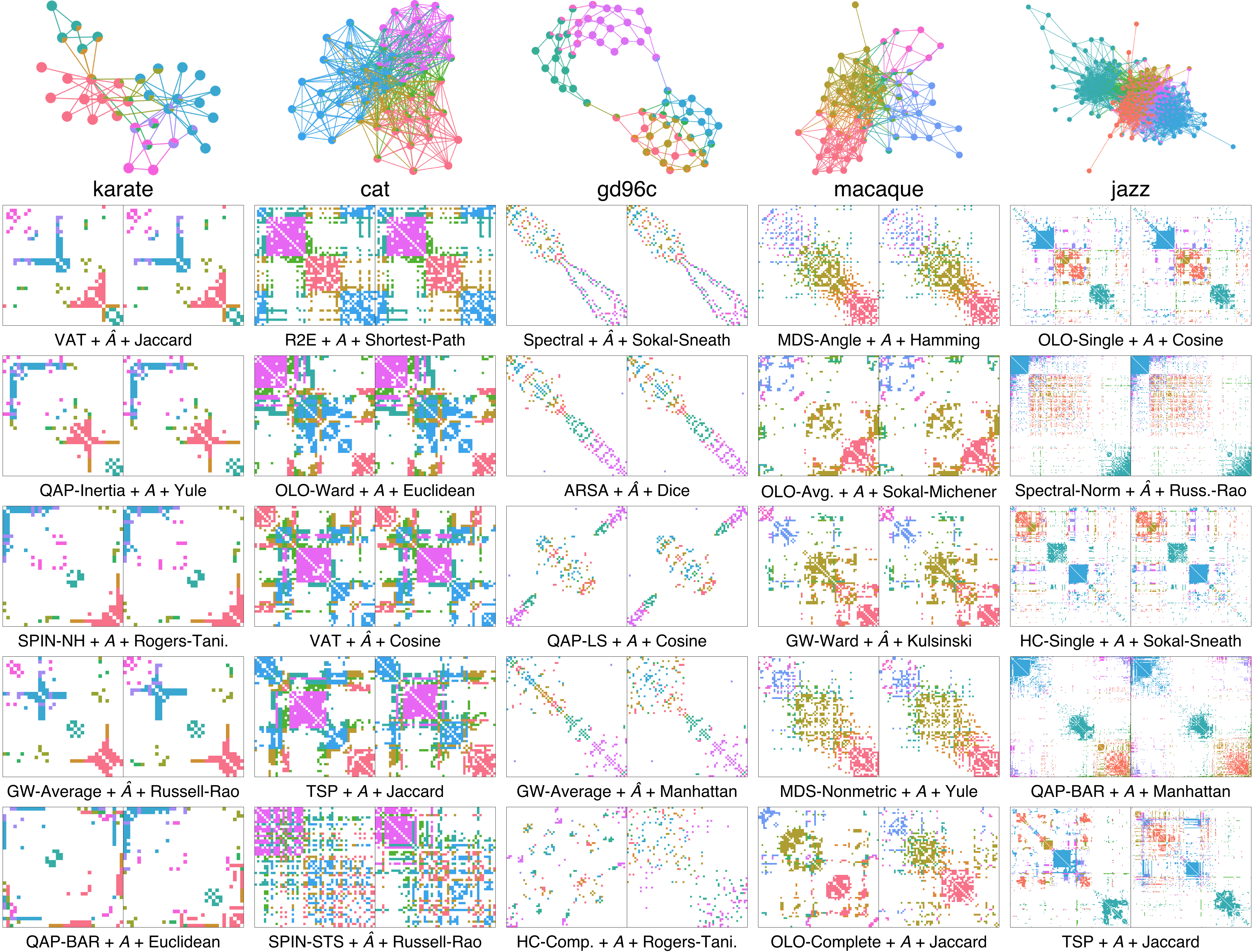}
\vspace{-1em}
\caption{%
Test reconstruction results of the five graphs using the MLP-Sinkhorn models.
For each pair of matrix reorderings, the left is test input, and the right is the reconstructed matrix reordering.
The label of a pair describes how the input matrix reordering is created: matrix reordering method (\autoref{dmr:tab:reordering_methods}), the type of the adjacency matrix ($A$: the original adjacency matrix, $\hat{A}$: the adjacency matrix with self-loop), and the distance metric (\autoref{dmr:tab:distances}).
The results are discussed in detail in \autoref{dmr:sec:eval:qualitative}.
The color of the matrices and node-link diagrams represents the community structure of the graph \cite{leiden}, so the readers can easily compare different matrix reorderings.
}
\label{dmr:fig:mlp-sinkhorn}
\end{figure*}

Overall, the MLP-Sinkhorn models show the lowest error rates for all the graphs (\autoref{dmr:sec:eval:quantitative}).
\autoref{dmr:fig:mlp-sinkhorn} shows several test reconstructions of five different graphs (\textsf{karate}, \textsf{cat}, \textsf{gd96c}, \textsf{macaque}, and \textsf{jazz}) using the MLP-Sinkhorn models.
The first three rows of reconstructions of \autoref{dmr:fig:mlp-sinkhorn} demonstrate that the MLP-Sinkhorn models can produce diverse styles of matrix reorderings of the given graph.
Especially, the reconstructions in the first row are exact reconstructions: the models produce unseen matrix reorderings without any difference.
Although exact test reconstructions do not guarantee that the model generates high-quality samples of a generative model, it does demonstrate the generalization capability of our models.

Further investigation of exact reconstructions also revealed that our architectures and loss function are indeed permutation equivariant (\autoref{dmr:sec:approach:perm-eq} and \autoref{dmr:sec:approach:architecture}).
For the \textsf{karate}, \textsf{collab}, \textsf{ego-fb}, and \textsf{jazz} graphs, 99.5\% of exact reconstruction cases have the generated permutation matrix $P'$ that is not the same as the permutation matrix of the input matrix reordering $P$.
However, this ratio depends on the number of automorphisms of a graph.
For other graphs, all the generated permutation matrix $P'$ is the same as the permutation matrix of the input matrix reordering $P$.

The bottom two rows of \autoref{dmr:fig:mlp-sinkhorn} show a number of reconstructions with higher losses (i.e., higher error rates).
However, the reconstructions in the fourth row are perceptually similar, although they have relatively higher losses than the first three rows.
These results show a limitation of our loss function: the per-cell loss can disagree with the human's perceptual similarity, which is a known issue in other computer vision tasks \cite{Zhang2018Perceptual}.
We believe a perceptual loss function \cite{Zhang2018Perceptual} can be used for improving the quality of the generated samples.
\section{Usage Scenario}
\begin{figure*}[t]
\captionsetup{farskip=0pt,skip=0pt}
\centering
\includegraphics[width=\textwidth, keepaspectratio]{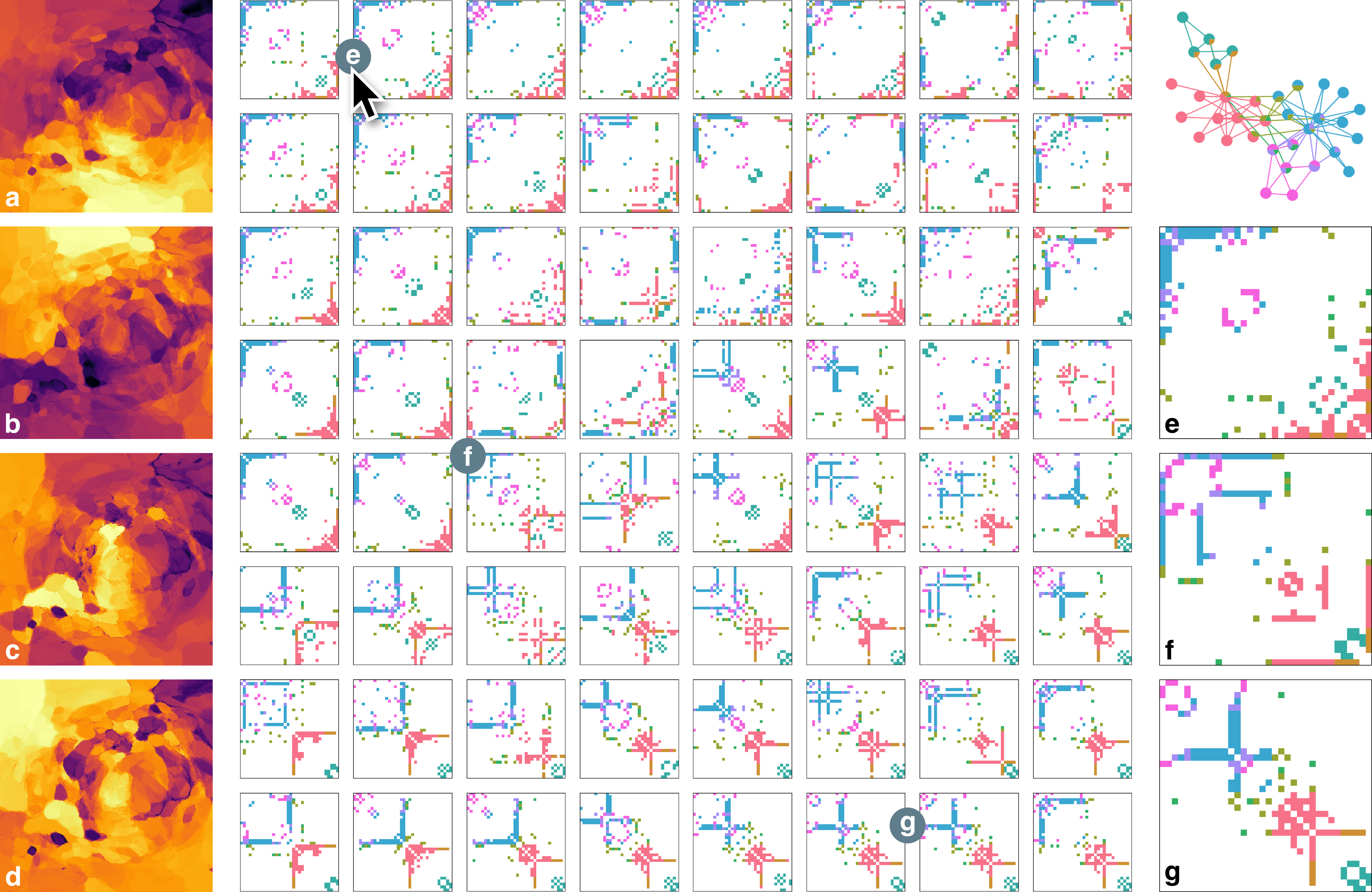}
\vspace{-1em}
\caption{%
The latent space of matrix reorderings for the \textsf{karate} graph learned by a MLP-Sinkhorn model.
The grid of samples (center) is generated by decoding $8 \times 8$ latent variables $z$ in $[-1, 1]^2$.
Users can intuitively explore diverse matrix visualizations and select one they want by simply pointing to a location in the latent space (e--g).
The heatmaps (a--d) show four quality metrics of $256 \times 256$ matrix reorderings:
(a)~$A$ (the original adjacency matrix) + Shortest-path distance (distance measure) + the number of anti-Robinson events (AR),
(b)~$A$ + Hamming + AR,
(c)~$A$ + Euclidean + Correlation,
and (d)~$\hat{A}$ (the adjacency matrix with self-loops) + Euclidean + Correlation.
The brighter colors represent ``better'' values in terms of the corresponding quality metric \cite{RSeriation}.
See \autoref{dmr:sec:eval:qualitative} for a detailed discussion.}
\label{dmr:fig:latent-space:karate}
\end{figure*}

\begin{figure*}[t]
\captionsetup{farskip=0pt,skip=0pt}
\centering
\includegraphics[width=\textwidth, keepaspectratio]{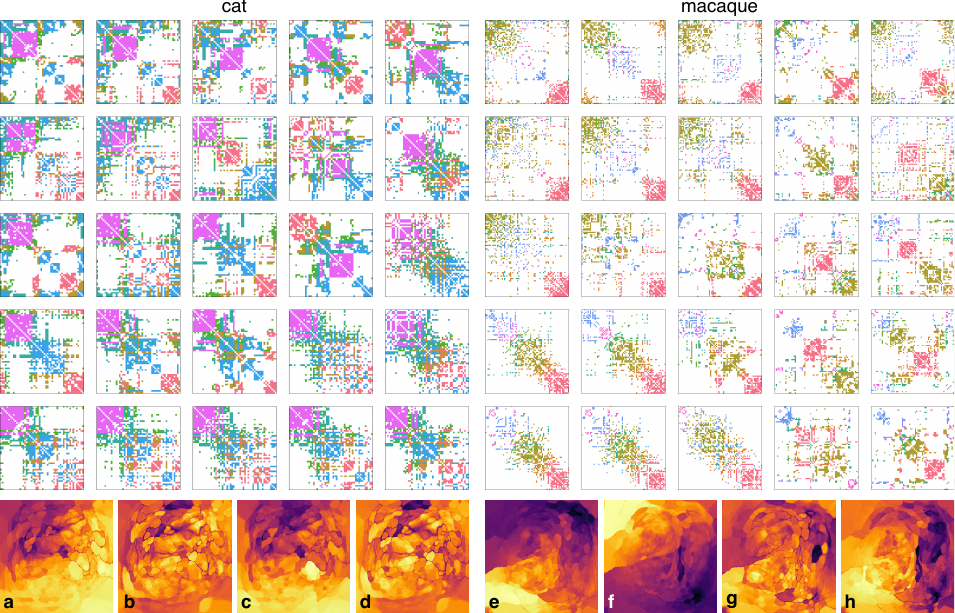}
\vspace{-1em}
\caption{%
The latent spaces of matrix reorderings for the \textsf{cat} and \textsf{macaque} graphs learned by MLP-Sinkhorn models.
The sample grids are generated by decoding $5 \times 5$ latent variables $z$ in $[-1, 1]^2$.
The same color scheme is used for the matrices as \autoref{dmr:fig:mlp-sinkhorn}.
The heatmaps (a--h) show several quality metrics of $256 \times 256$ matrix reorderings in $[-1, 1]^2$:
(a)~$A$ (the original adjacency matrix) + Shortest-path distance (distance measure) + the number of anti-Robinson events (AR),
(b)~$A$ + Shortest-path distance + the closeness to the banded anti-Robinson form (BAR),
(c)~$A$ + Yule + AR,
(d)~$A$ + Yule + BAR,
(e)~$A$ + Shortest-path distance + AR,
(f)~$A$ + Hamming + AR,
(g)~$\hat{A}$ (the adjacency matrix with self-loops) + Cosine + BAR,
and (h)~$A$ + Yule + Correlation.
}
\label{dmr:fig:latent-space:cat-macaque}
\end{figure*}

We discuss how a learned latent space can support users in designing effective matrix visualizations of the given graph.
Browsing multiple matrix reorderings and selecting the desired one from an unsorted list often lead to tedious trials and errors.
Since we train a model to construct a 2D latent space, a grid of generated samples can be created to show several representative matrix reorderings of the given graph that the model learned.
\autoref{dmr:fig:latent-space:karate} and \autoref{dmr:fig:latent-space:cat-macaque} show the latent spaces of matrix reorderings for the \textsf{karate}, \textsf{cat}, and \textsf{macaque} graphs learned by MLP-Sinkhorn models.
The sample grids provide an organized overview of diverse matrix reorderings of the given graph, where similar matrix reorderings are placed close to each other in the latent space.

Using a sample grid as a \emph{map} of the latent space of diverse matrix reorderings,
we build an intuitive interface, where users can directly set the latent variable $z$ to produce different matrix reorderings and select one they want by simply pointing to a location in the latent space.
For example, \autoref{dmr:fig:latent-space:karate} shows a grid of $8 \times 8$ samples throughout the latent space in $[-1, 1]^2$, where the bottom left-corner is $(-1, -1)$ and the top-right corner is $(1, 1)$.

When the user points the location (\textsf{e}) in the latent space, the latent variable $z$ is set to $(-0.746, 0.873)$.
Then, the model produces the corresponding reordering shown in \autoref{dmr:fig:latent-space:karate}e on the right.
Users can effortlessly produce different matrix reorderings by pointing to different locations.
Pointing the location (\textsf{f}) generates a matrix reordering that highlights the highly-connected nodes of the graph: the \colorbox[rgb]{0.224, 0.655, 0.816}{blue} and \colorbox[rgb]{0.969, 0.443, 0.537}{red} cells that form lines in \autoref{dmr:fig:latent-space:karate}f on the right.
From the location (\textsf{g}) in the latent space, users can produce a matrix reordering that can show a pattern that is not obvious in the other reorderings (\textsf{e} and \textsf{f}): the \colorbox[rgb]{0.592156863, 0.643137255, 0.192156863}{olive} cells are the edges that connect the \colorbox[rgb]{0.224, 0.655, 0.816}{blue} and \colorbox[rgb]{0.969, 0.443, 0.537}{red} clusters.

Since a learned latent space is $\mathbb{R}^2$, we can create sample grids with different levels of granularity.
For instance, a grid of a smaller number of samples can be used for showing more details of the generated adjacency matrices of a larger graph in a limited display area, as shown in \autoref{dmr:fig:latent-space:cat-macaque}.
In addition, we can visualize a latent space as a heatmap with a large number of samples to support experts in matrix reordering (e.g., algorithm designer).

Multiple quality metrics have been introduced to measure the ``goodness'' of a matrix reordering \cite{RSeriation}.
Moreover, we can define different distance measures between the nodes of a graph (\autoref{dmr:sec:approach:training_data}).
Thus, finding the quality metric and distance measure that can quantify the desired quality of a matrix reordering for the given graph and analysis goal is not a trivial task.

\autoref{dmr:fig:latent-space:karate}a--d and \autoref{dmr:fig:latent-space:cat-macaque}a--h show several heatmaps that represent quality metrics of $256 \times 256$ matrix reorderings of the corresponding graph, where the samples from the brighter areas have ``better'' values in terms of the definition of each metric (e.g., lower loss values or higher merit values \cite{RSeriation}).
For example, \autoref{dmr:fig:latent-space:karate}a is a map of the number
of anti-Robinson events \cite{R2E, Tien08} using the shortest-path distances between nodes, where the areas with brighter colors have a smaller number of anti-Robinson events.
This metric could find a matrix reordering like \autoref{dmr:fig:latent-space:karate}g, which can better reveal the \colorbox[rgb]{0.592156863, 0.643137255, 0.192156863}{olive} edges between the \colorbox[rgb]{0.224, 0.655, 0.816}{blue} and \colorbox[rgb]{0.969, 0.443, 0.537}{red} clusters.

By visualizing the latent space with quality metrics, experts in matrix reordering can understand what patterns the given metric can capture or not with diverse and concrete samples.
For example, we can clearly see the differences between
(1) different quality metrics (\autoref{dmr:fig:latent-space:cat-macaque}a and b, and \autoref{dmr:fig:latent-space:cat-macaque}c and d),
(2) different distance measures (\autoref{dmr:fig:latent-space:karate}a and b, and \autoref{dmr:fig:latent-space:karate}c and d), and
(3) different combinations of quality metrics and distance measures (\autoref{dmr:fig:latent-space:cat-macaque}e--h).
Thus, these heatmaps can support the process of designing a new quality metric or distance measure.

In summary, the learned latent spaces support users in designing effective matrix visualizations of a graph by providing a map of diverse matrix reorderings.
A further evaluation similar to the user study of \cite{Kwon18} could quantify the quality of the latent space by comparing the proximity in the latent space and the perceptual similarity of generated matrix reorderings.
\section{Conclusion}
Representing a graph as an adjacency matrix is a popular approach to graph visualization in addition to using a node-link diagram.
However, different node orderings can reveal varying structural characteristics of the same graph.
Therefore, finding a ``good'' node ordering is an important step for visualizing a graph as adjacency matrix views.
Users generally rely on a trial-and-error effort to find a good node ordering.

In this paper, we have introduced a deep generative model that can learn a latent space of matrix reorderings of the given graph.
The learned latent space is used as a WYSIWYG interface, where users can effortlessly explore diverse matrix reorderings and select the desired one for the purpose of visualization.
To achieve this, we have designed a novel neural network architecture to address unique challenges in learning a deep generative model for matrix reordering.

A graph can be visualized in many different ways, where each visualization can highlight different characteristics of the same graph, leading to new insights and discoveries.
We hope this paper encourages others to join a new area of graph visualization study, where the goal is to design a method that produces diverse visualizations of the same graph, not just a single visualization that follows only a certain set of heuristics.


%



\ifCLASSOPTIONcompsoc
  \section*{Acknowledgments}
\else
  \section*{Acknowledgment}
\fi

This research is sponsored in part by the U.S. National Science Foundation
through grant IIS-1741536. 
\vfill

\ifCLASSOPTIONcaptionsoff
  \newpage
\fi



\bibliographystyle{IEEEtran}
\bibliography{tex/references}

\begin{thebibliography}{10}
\providecommand{\url}[1]{#1}
\csname url@samestyle\endcsname
\providecommand{\newblock}{\relax}
\providecommand{\bibinfo}[2]{#2}
\providecommand{\BIBentrySTDinterwordspacing}{\spaceskip=0pt\relax}
\providecommand{\BIBentryALTinterwordstretchfactor}{4}
\providecommand{\BIBentryALTinterwordspacing}{\spaceskip=\fontdimen2\font plus
\BIBentryALTinterwordstretchfactor\fontdimen3\font minus
  \fontdimen4\font\relax}
\providecommand{\BIBforeignlanguage}[2]{{%
\expandafter\ifx\csname l@#1\endcsname\relax
\typeout{** WARNING: IEEEtran.bst: No hyphenation pattern has been}%
\typeout{** loaded for the language `#1'. Using the pattern for}%
\typeout{** the default language instead.}%
\else
\language=\csname l@#1\endcsname
\fi
#2}}
\providecommand{\BIBdecl}{\relax}
\BIBdecl

\bibitem{Munzner14:VAD}
T.~Munzner, \emph{{Visualization Analysis and Design}}.\hskip 1em plus 0.5em
  minus 0.4em\relax {CRC Press}, 2014.

\bibitem{WuTzengChen08}
H.-M. Wu, S.~Tzeng, and C.-h. Chen, ``{Matrix Visualization},'' in
  \emph{{Handbook of Data Visualization}}, C.-h. Chen, W.~H\"{a}rdle, and
  A.~Unwin, Eds.\hskip 1em plus 0.5em minus 0.4em\relax Springer, 2008, pp.
  681--708.

\bibitem{Liiv10}
I.~Liiv, ``{Seriation and Matrix Reordering Methods: An Historical Overview},''
  \emph{Statistical Analysis and Data Mining}, vol.~3, no.~2, pp. 70--91, 2010.

\bibitem{Behrisch16}
M.~Behrisch, B.~Bach, N.~Henry~Riche, T.~Schreck, and J.-D. Fekete, ``{Matrix
  Reordering Methods for Table and Network Visualization},'' \emph{Computer
  Graphics Forum}, vol.~35, no.~3, pp. 693--716, 2016.

\bibitem{Kwon19}
O.-H. Kwon and K.-L. Ma, ``{A Deep Generative Model for Graph Layout},''
  \emph{IEEE Transactions on Visualization and Computer Graphics}, vol.~26,
  no.~1, pp. 665--675, 2020.

\bibitem{DeepLearningNature}
Y.~LeCun, Y.~Bengio, and G.~Hinton, ``{Deep Learning},'' \emph{Nature}, vol.
  521, no. 7553, pp. 436--444, 2015.

\bibitem{Sinkhorn}
G.~Mena, D.~Belanger, S.~Linderman, and J.~Snoek, ``{Learning Latent
  Permutations with Gumbel-Sinkhorn Networks},'' in \emph{Proc. International
  Conference on Learning Representations}, 2018.

\bibitem{SoftSort}
S.~Prillo and J.~M. Eisenschlos, ``{SoftSort: A Continuous Relaxation for the
  argsort Operator},'' in \emph{Proc. International Conference on Machine
  Learning}, 2020, pp. 7793--7802.

\bibitem{TSPOrdering}
E.~Wilkinson, ``{Archaeological Seriation and the Travelling Salesman
  Problem},'' in \emph{{Mathematics in the Archaeological and Historical
  Sciences}}, F.~Hodson, D.~Kendall, and P.~Tautu, Eds.\hskip 1em plus 0.5em
  minus 0.4em\relax Edinburgh University Press, 1971, pp. 276--284.

\bibitem{SpectralOrdering}
S.~T. {Barnard}, A.~{Pothen}, and H.~D. {Simon}, ``{A Spectral Algorithm for
  Envelope Reduction of Sparse Matrices},'' in \emph{Proc. ACM/IEEE Conference
  on Supercomputing}, 1993, pp. 493--502.

\bibitem{DingHe04}
C.~Ding and X.~He, ``{Linearized Cluster Assignment via Spectral Ordering},''
  in \emph{Proc. International Conference on Machine Learning}, 2004.

\bibitem{Eisen98}
M.~B. Eisen, P.~T. Spellman, P.~O. Brown, and D.~Botstein, ``{Cluster Analysis
  and Display of Genome-Wide Expression Patterns},'' \emph{Proc. National
  Academy of Sciences}, vol.~95, no.~25, pp. 14\,863--14\,868, 1998.

\bibitem{OLO}
Z.~Bar-Joseph, D.~K. Gifford, and T.~S. Jaakkola, ``{Fast Optimal Leaf Ordering
  for Hierarchical Clustering},'' \emph{Bioinformatics}, vol.~17, no. Suppl. 1,
  pp. 22--29, 2001.

\bibitem{Gruvaeus72}
G.~Gruvaeus and H.~Wainer, ``{Two Additions to Hierarchical Cluster
  Analysis},'' \emph{British Journal of Mathematical and Statistical
  Psychology}, vol.~25, no.~2, pp. 200--206, 1972.

\bibitem{Hahsler17}
M.~Hahsler, ``{An Experimental Comparison of Seriation Methods for One-Mode
  Two-Way Data},'' \emph{European Journal of Operational Research}, vol. 257,
  no.~1, pp. 133--143, 2017.

\bibitem{Behrisch17}
M.~Behrisch, B.~Bach, M.~Hund, M.~Delz, L.~V. R\"{u}den, J.~D. Fekete, and
  T.~Schreck, ``{Magnostics: Image-Based Search of Interesting Matrix Views for
  Guided Network Exploration},'' \emph{IEEE Transactions on Visualization and
  Computer Graphics}, vol.~23, no.~1, pp. 31--40, 2017.

\bibitem{GUIRO}
M.~{Behrisch}, T.~{Schreck}, and H.~{Pfister}, ``{GUIRO: User-Guided Matrix
  Reordering},'' \emph{{IEEE Transactions on Visualization and Computer
  Graphics}}, vol.~26, no.~1, pp. 184--194, 2020.

\bibitem{MatrixExplorer}
N.~{Henry} and J.~{Fekete}, ``{MatrixExplorer: A Dual-Representation System to
  Explore Social Networks},'' \emph{IEEE Transactions on Visualization and
  Computer Graphics}, vol.~12, no.~5, pp. 677--684, 2006.

\bibitem{PermutMatrix}
G.~Caraux and S.~Pinloche, ``{PermutMatrix: A Graphical Environment to Arrange
  Gene Expression Profiles in Optimal Linear Order},'' \emph{Bioinformatics},
  vol.~21, no.~7, pp. 1280--1281, 2004.

\bibitem{COMBat}
R.~B.~J. {van Brakel}, M.~W. E.~J. {Fiers}, C.~{Francke}, M.~A. {Westenberg},
  and H.~{van de Wetering}, ``{COMBat: Visualizing Co-occurrence of Annotation
  Terms},'' in \emph{Proc. IEEE Symposium on Biological Data Visualization},
  2013, pp. 17--24.

\bibitem{VAE}
D.~P. Kingma and M.~Welling, ``{Auto-Encoding Variational Bayes},'' in
  \emph{Proc. International Conference on Learning Representations}, 2014.

\bibitem{GAN}
I.~Goodfellow, J.~Pouget-Abadie, M.~Mirza, B.~Xu, D.~Warde-Farley, S.~Ozair,
  A.~Courville, and Y.~Bengio, ``{Generative Adversarial Nets},'' in
  \emph{Proc. Advances in Neural Information Processing Systems}, 2014, pp.
  2672--2680.

\bibitem{VQVAE2}
\BIBentryALTinterwordspacing
A.~Razavi, A.~van~den Oord, and O.~Vinyals, ``{Generating Diverse High-Fidelity
  Images with VQ-VAE-2},'' \emph{arXiv preprint}, vol. arXiv:1906.00446, 2019.
  [Online]. Available: \url{https://arxiv.org/abs/1906.00446}
\BIBentrySTDinterwordspacing

\bibitem{BigGAN}
\BIBentryALTinterwordspacing
A.~Brock, J.~Donahue, and K.~Simonyan, ``{Large Scale GAN Training for High
  Fidelity Natural Image Synthesis},'' \emph{arXiv preprint}, vol.
  arXiv:1809.11096, 2018. [Online]. Available:
  \url{https://arxiv.org/abs/1809.11096}
\BIBentrySTDinterwordspacing

\bibitem{Guu18TextGen}
K.~Guu, T.~B. Hashimoto, Y.~Oren, and P.~Liang, ``Generating sentences by
  editing prototypes,'' \emph{Transactions of the Association for Computational
  Linguistics}, vol.~6, pp. 437--450, 2018.

\bibitem{Zhang17TextGen}
Y.~Zhang, Z.~Gan, K.~Fan, Z.~Chen, R.~Henao, D.~Shen, and L.~Carin,
  ``{Adversarial Feature Matching for Text Generation},'' in \emph{Proc.
  International Conference on Machine Learning}, 2017, p. 4006–4015.

\bibitem{Jukebox}
\BIBentryALTinterwordspacing
P.~Dhariwal, H.~Jun, C.~Payne, J.~W. Kim, A.~Radford, and I.~Sutskever,
  ``{Jukebox: A Generative Model for Music},'' \emph{arXiv preprint}, vol.
  arXiv:2005.00341, 2020. [Online]. Available:
  \url{https://arxiv.org/abs/2005.00341}
\BIBentrySTDinterwordspacing

\bibitem{GANSynth}
J.~Engel, K.~K. Agrawal, S.~Chen, I.~Gulrajani, C.~Donahue, and A.~Roberts,
  ``{GANSynth: Adversarial Neural Audio Synthesis},'' in \emph{Proc.
  International Conference on Learning Representations}, 2019.

\bibitem{NeuralSort}
A.~Grover, E.~Wang, A.~Zweig, and S.~Ermon, ``{Stochastic Optimization of
  Sorting Networks via Continuous Relaxations},'' in \emph{Proc. International
  Conference on Learning Representations}, 2019.

\bibitem{Hungarian}
H.~W. Kuhn, ``{The Hungarian Method for the Assignment Problem},'' \emph{Naval
  Research Logistics Quarterly}, vol.~2, no. 1--2, pp. 83--97, 1955.

\bibitem{Climer06}
S.~Climer and W.~Zhang, ``{Rearrangement Clustering: Pitfalls, Remedies, and
  Applications},'' \emph{Journal of Machine Learning Research}, vol.~7, pp.
  919--943, 2006.

\bibitem{Brusco08}
M.~J. Brusco, H.-F. K{\"o}hn, and S.~Stahl, ``{Heuristic Implementation of
  Dynamic Programming for Matrix Permutation Problems in Combinatorial Data
  Analysis},'' \emph{Psychometrika}, vol.~73, no.~3, pp. 503--522, 2008.

\bibitem{R2E}
C.-h. Chen, ``{Generalized Association Plots: Information Visualization via
  Iteratively Generated Correlation Matrices},'' \emph{Statistica Sinica},
  vol.~12, pp. 7--29, 2002.

\bibitem{DCGAN}
A.~Radford, L.~Metz, and S.~Chintala, ``{Unsupervised Representation Learning
  with Deep Convolutional Generative Adversarial Networks},'' \emph{arXiv
  preprint}, vol. arXiv:1511.06434, 2015.

\bibitem{SWAE}
S.~Kolouri, P.~E. Pope, C.~E. Martin, and G.~K. Rohde, ``{Sliced-Wasserstein
  Auto-Encoders},'' in \emph{Proc. International Conference on Learning
  Representations}, 2019.

\bibitem{konect}
\BIBentryALTinterwordspacing
J.~Kunegis, ``{KONECT} -- {The} {Koblenz} {Network} {Collection},'' in
  \emph{Proc. International Conference on World Wide Web Companion}, 2013, pp.
  1343--1350. [Online]. Available:
  \url{http://dl.acm.org/citation.cfm?id=2488173}
\BIBentrySTDinterwordspacing

\bibitem{Karate}
W.~W. Zachary, ``{An Information Flow Model for Conflict and Fission in Small
  Groups},'' \emph{Journal of Anthropological Research}, vol.~30, pp. 452--473,
  1977.

\bibitem{CatCortex}
J.~W. Scannell, G.~A. Burns, C.~C. Hilgetag, M.~A. O'Neil, and M.~P. Young,
  ``{The Connectional Organization of the Cortico-Thalamic System of the
  Cat},'' \emph{Cerebral Cortex}, vol.~9, no.~3, pp. 277--299, 1999.

\bibitem{GD96C}
J.~Petit, ``{Experiments on the Minimum Linear Arrangement Problem},''
  \emph{ACM Journal of Experimental Algorithmics}, vol.~8, p. 2.3, 2004.

\bibitem{Young93}
M.~P. Young, ``{The Organization of Neural Systems in the Primate Cerebral
  Cortex},'' \emph{Proc. Royal Society of London B: Biological Sciences}, vol.
  252, no. 1333, pp. 13--18, 1993.

\bibitem{Yanardag15}
P.~Yanardag and S.~Vishwanathan, ``{Deep Graph Kernels},'' in \emph{Proc. ACM
  SIGKDD Conference on Knowledge Discovery and Data Mining}, 2015, pp.
  1365--1374.

\bibitem{SNAP}
J.~Leskovec and A.~Krevl, ``{SNAP Datasets: Stanford Large Network Dataset
  Collection},'' \url{http://snap.stanford.edu/data}, Jun. 2014.

\bibitem{EgoFacebook}
J.~McAuley and J.~Leskovec, ``{Learning to Discover Social Circles in Ego
  Networks},'' in \emph{Proc. Advances in Neural Information Processing
  Systems}, 2012, pp. 539--547.

\bibitem{Jazz}
P.~M. Gleiser and D.~Leon, ``{Community Structure in Jazz},'' \emph{Advances in
  Complex Systems}, vol.~6, no.~4, pp. 565--573, 2003.

\bibitem{RSeriation}
M.~Hahsler, K.~Hornik, and C.~Buchta, ``{Getting Things in Order: An
  Introduction to the R Package seriation},'' \emph{Journal of Statistical
  Software, Articles}, vol.~25, no.~3, pp. 1--34, 2008.

\bibitem{MDSKendall}
F.~Hodson, D.~Kendall, and P.~Tautu, ``{Seriation from Abundance Matrices},''
  in \emph{{Mathematics in the Archaeological and Historical Sciences}}.\hskip
  1em plus 0.5em minus 0.4em\relax Edinburgh University Press, 1971, pp.
  215--252.

\bibitem{VAT}
J.~C. {Bezdek} and R.~J. {Hathaway}, ``{VAT: A Tool for Visual Assessment of
  (Cluster) Tendency},'' in \emph{Proc. International Joint Conference on
  Neural Networks}, vol.~3, 2002, pp. 2225--2230.

\bibitem{Tsafrir05}
D.~Tsafrir, I.~Tsafrir, L.~Ein-Dor, O.~Zuk, D.~Notterman, and E.~Domany,
  ``{Sorting Points into Neighborhoods (SPIN): Data Analysis and Visualization
  by Ordering Distance Matrices},'' \emph{Bioinformatics}, vol.~21, no.~10, pp.
  2301--2308, 2005.

\bibitem{HubertSchultz76}
L.~Hubert and J.~Schultz, ``{Quadratic Assignment as a General Data Analysis
  Strategy},'' \emph{British Journal of Mathematical and Statistical
  Psychology}, vol.~29, no.~2, pp. 190--241, 1976.

\bibitem{BandedAntiRobinson}
D.~Earle and C.~B. Hurley, ``{Advances in Dendrogram Seriation for Application
  to Visualization},'' \emph{Journal of Computational and Graphical
  Statistics}, vol.~24, no.~1, pp. 1--25, 2015.

\bibitem{LayerNorm}
J.~L. Ba, J.~R. Kiros, and G.~E. Hinton, ``{Layer Normalization},'' \emph{arXiv
  preprint}, vol. arXiv:1607.06450, 2016.

\bibitem{ELU}
D.-A. Clevert, T.~Unterthiner, and S.~Hochreiter, ``{Fast and Accurate Deep
  Network Learning by Exponential Linear Units (ELUs)},'' in \emph{Proc.
  International Conference on Learning Representations}, 2016.

\bibitem{BatchNorm}
S.~Ioffe and C.~Szegedy, ``{Batch Normalization: Accelerating Deep Network
  Training by Reducing Internal Covariate Shift},'' in \emph{Proc.
  International Conference on Machine Learning}, 2015, pp. 448--456.

\bibitem{Adam}
S.~J. Reddi, S.~Kale, and S.~Kumar, ``{On the Convergence of Adam and
  Beyond},'' in \emph{Proc. International Conference on Learning
  Representations}, 2018.

\bibitem{PyTorch}
A.~Paszke, S.~Gross, S.~Chintala, G.~Chanan, E.~Yang, Z.~DeVito, Z.~Lin,
  A.~Desmaison, L.~Antiga, and A.~Lerer, ``{Automatic Differentiation in
  PyTorch},'' in \emph{Proc. Advances in Neural Information Processing Systems
  2017 Autodiff Workshop}, 2017.

\bibitem{StyleGAN2}
T.~Karras, S.~Laine, M.~Aittala, J.~Hellsten, J.~Lehtinen, and T.~Aila,
  ``{Analyzing and Improving the Image Quality of StyleGAN},'' in \emph{Proc.
  IEEE Conference on Computer Vision and Pattern Recognition}, 2020, pp.
  8107--8116.

\bibitem{VAinDL}
F.~Hohman, M.~Kahng, R.~Pienta, and D.~H. Chau, ``{Visual Analytics in Deep
  Learning: An Interrogative Survey for the Next Frontiers},'' \emph{IEEE
  Transactions on Visualization and Computer Graphics}, vol.~25, no.~8, pp.
  2674--2693, 2019.

\bibitem{NonLocal}
X.~{Wang}, R.~{Girshick}, A.~{Gupta}, and K.~{He}, ``{Non-local Neural
  Networks},'' in \emph{Proc. IEEE/CVF Conference on Computer Vision and
  Pattern Recognition}, 2018, pp. 7794--7803.

\bibitem{SAGAN}
H.~Zhang, I.~Goodfellow, D.~Metaxas, and A.~Odena, ``{Self-Attention Generative
  Adversarial Networks},'' in \emph{Proc. International Conference on Machine
  Learning}, 2019, pp. 7354--7363.

\bibitem{StdAlnSelfAttn}
P.~Ramachandran, N.~Parmar, A.~Vaswani, I.~Bello, A.~Levskaya, and J.~Shlens,
  ``{Stand-Alone Self-Attention in Vision Models},'' in \emph{Advances in
  Neural Information Processing Systems}, 2019, pp. 68--80.

\bibitem{GCNet}
Y.~Cao, J.~Xu, S.~Lin, F.~Wei, and H.~Hu, ``{GCNet: Non-local Networks Meet
  Squeeze-Excitation Networks and Beyond},'' in \emph{Proc. IEEE/CVF
  International Conference on Computer Vision Workshop}, 2019, pp. 1971--1980.

\bibitem{EffAttn}
\BIBentryALTinterwordspacing
Z.~Shen, M.~Zhang, S.~Yi, J.~Yan, and H.~Zhao, ``{Efficient Attention:
  Attention with Linear Complexities},'' \emph{arXiv preprint}, vol.
  arXiv:1812.01243, 2018. [Online]. Available:
  \url{https://arxiv.org/abs/1812.01243}
\BIBentrySTDinterwordspacing

\bibitem{Barabasi16}
A.-L. Barab{\'a}si, \emph{{Network Science}}.\hskip 1em plus 0.5em minus
  0.4em\relax Cambridge University Press, 2016.

\bibitem{leiden}
V.~A. Traag, L.~Waltman, and N.~J. van Eck, ``{From Louvain to Leiden:
  Guaranteeing Well-Connected Communities},'' \emph{Scientific Reports},
  vol.~9, no.~1, p. 5233, 2019.

\bibitem{Zhang2018Perceptual}
R.~Zhang, P.~Isola, A.~A. Efros, E.~Shechtman, and O.~Wang, ``{The Unreasonable
  Effectiveness of Deep Features as a Perceptual Metric},'' in \emph{Proc. IEEE
  Conference on Computer Vision and Pattern Recognition}, 2018, pp. 586--595.

\bibitem{Tien08}
Y.-J. Tien, Y.-S. Lee, H.-M. Wu, and C.-h. Chen, ``{Methods for Simultaneously
  Identifying Coherent Local Clusters with Smooth Global Patterns in Gene
  Expression Profiles},'' \emph{BMC Bioinformatics}, vol.~9, no. 155, 2008.

\bibitem{Kwon18}
O.-H. Kwon, T.~Crnovrsanin, and K.-L. Ma, ``{What Would a Graph Look Like in
  This Layout? A Machine Learning Approach to Large Graph Visualization},''
  \emph{IEEE Transactions on Visualization and Computer Graphics}, vol.~24,
  no.~1, pp. 478--488, 2018.

\end{thebibliography}
%



\vfill\eject

%
\begin{IEEEbiography}[{\includegraphics[width=1in,height=1.25in,clip,keepaspectratio]{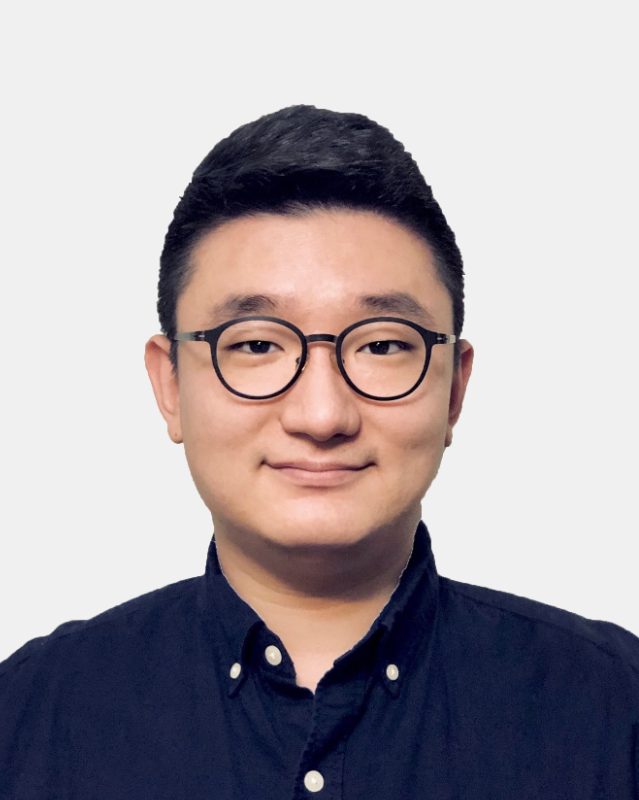}}]{Oh-Hyun Kwon} received the PhD degree in computer science from the University of California, Davis in 2021.
His PhD research focused on developing machine learning and immersive approaches to graph visualization.
He is currently working in the industry in the areas of data visualization, visual analytics, and machine learning.
\end{IEEEbiography}

\begin{IEEEbiography}[{\includegraphics[width=1in,height=1.25in,clip,keepaspectratio]{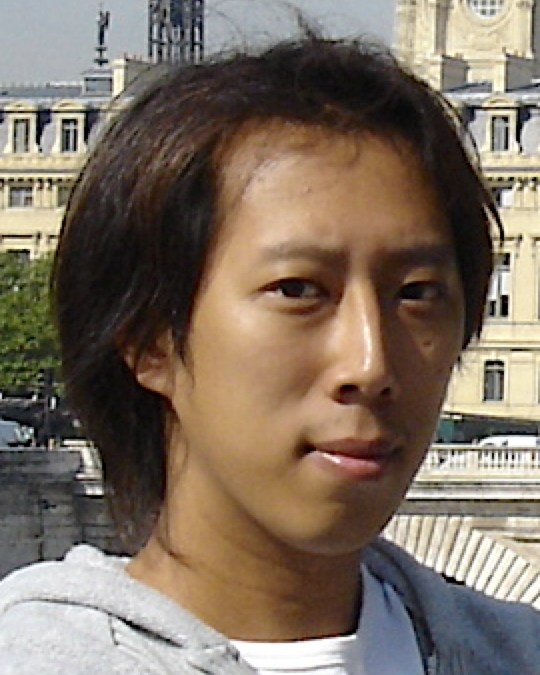}}]{Chiun-How Kao}
received the PhD degree in information management from the National Taiwan University of Science and Technology in 2018. 
He is a assistant professor in the Department of Statistics at Tamkang University. 
His research interests include data visualization, symbolic data analysis, and multimedia systems.
\end{IEEEbiography}

\begin{IEEEbiography}[{\includegraphics[width=1in,height=1.25in,clip,keepaspectratio]{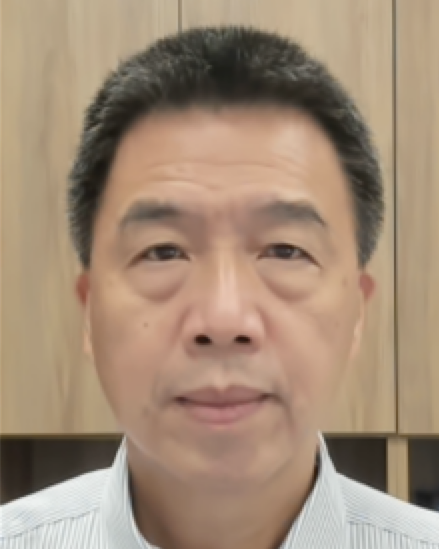}}]{Chun-houh Chen} received the PhD degree in Mathematics (Program in Statistics) from the University of California, Los Angeles in 1992. 
He is a research fellow and director of the Institute of Statistical Science, Academia Sinica in Taiwan. 
His research interests include matrix visualization, biobanking and smart health, bioinformatics, dimensionality reduction, high-dimensional statistical methods. 
He is now president-elect of the International Association for Statistical Computing.
\end{IEEEbiography}

\begin{IEEEbiography}[{\includegraphics[width=1in,height=1.25in,clip,keepaspectratio]{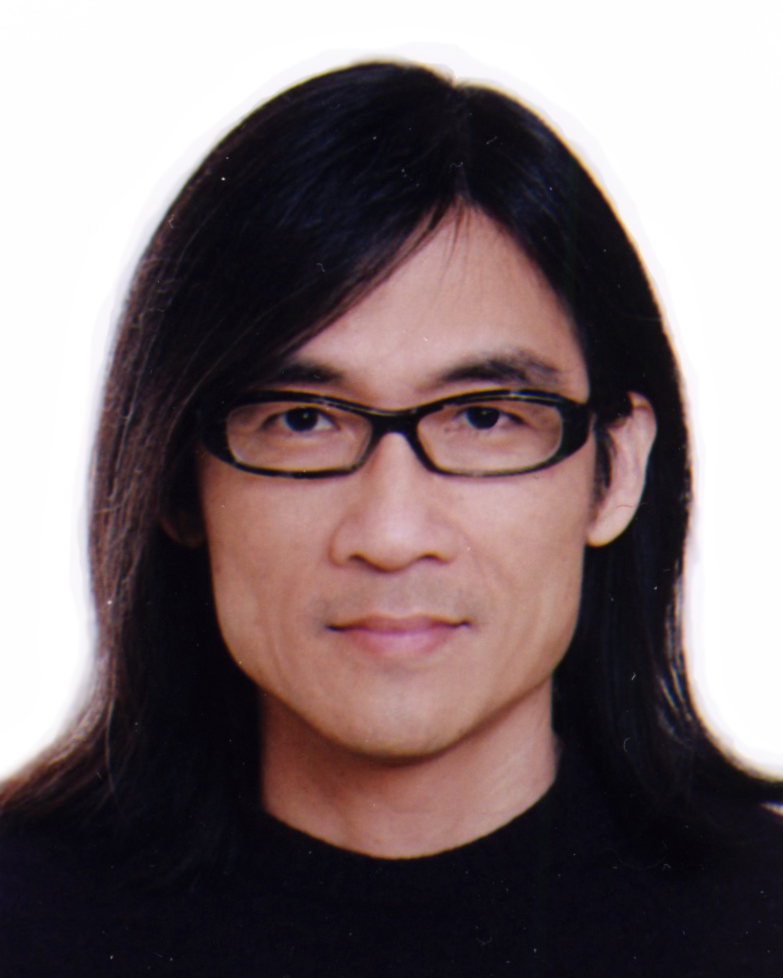}}]{Kwan-Liu Ma} is a distinguished professor of computer science at the University of California, Davis. 
He received his PhD degree in computer science from the University of Utah in 1993, and then worked as a staff scientist at ICASE/NASA Langley Research Center before joining UC Davis in 1999. 
His research is in the intersection of data visualization, computer graphics, human-computer interaction, and high performance computing. 
For his significant research accomplishments, Ma received international recognition, which includes being elected as IEEE Fellow in 2012, recipient of the IEEE VGTC Visualization Technical Achievement Award in 2013, and inducted to IEEE Visualization Academy in 2019.
\end{IEEEbiography}





\vfill


\end{document}